\documentclass[twocolumn,showpacs,preprintnumbers,amsmath,amssymb,superscriptaddress,aps,pre]{revtex4-1}

\usepackage{graphicx}
\usepackage{dcolumn}
\usepackage{bm}
\usepackage{subfigure}
\usepackage{color}

\begin{document}

\preprint{APS/123-QED}

\title{Lattice Boltzmann - Langevin simulations of binary mixtures}

\author{P.T. Sumesh}
\affiliation{Engineering Mechanics Unit, Jawaharlal Nehru Centre for Advanced Scientific Research, Bangalore 560064, India}
\author{Ignacio Pagonabarraga}
\affiliation{Department de Fisica Fonamental, Universitat de Barcelona, Avinguda Diagonal 647, E-08028 Barcelona, Spain}
\author{R. Adhikari} 
\affiliation{The Institute of Mathematical Sciences, CIT Campus, Chennai 600113, India}

\begin{abstract}
We report a hybrid numerical method for the solution of the model H fluctuating hydrodynamic equations for binary mixtures. The momentum conservation equations with Landau-Lifshitz stresses are solved using the fluctuating lattice Boltzmann equation while the order parameter conservation equation with Langevin fluxes are solved using the stochastic method of lines. Two methods, based on finite difference and finite volume, are proposed for spatial discretisation of the order parameter equation. Special care is taken to ensure that the fluctuation-dissipation theorem is maintained at the lattice level in both cases. The methods are benchmarked by comparing static and dynamic correlations and excellent agreement is found between analytical and numerical results. The Galilean invariance of the model is tested and found to be satisfactory. Thermally induced capillary fluctuations of the interface are captured accurately, indicating that the model can be used to study nonlinear fluctuations.
\end{abstract}

\pacs{Valid PACS appear here}
\maketitle

\section{\label{sec:level1}Introduction}

Thermal fluctuations are an essential part of the physics at mesoscopic length scales in fluid mechanical problems. For instance, thermal fluctuations produce Brownian motion in colloidal suspensions, conformational fluctuations of polymers and membranes, capillary waves at fluctuating interfaces, and critical opalescence in binary mixtures. A consistent mesoscopic description of such phenomena follows from the equations of fluctuating hydrodynamics. The first instance of such a description was the fluctuating Navier-Stokes equations of Landau and Lifshitz \cite{LandauBook}. Similar equations were then introduced to study the dynamics of order parameter fluctuations in critical phenomena, as reviewed by Halperin and Hohenberg \cite{halperin1977}. The coupled fluctuating equations of motion for the momentum and order parameter are known as model H in their classification. 

The model H equations describe the fluctuating hydrodynamics of a conserved order parameter $\psi$ and the conserved momentum density ${\bf g} = \rho {\bf u}$, where $\rho$ and ${\bf u}$ are the total density and the local fluid velocity. To ensure conservation of local densities, fluctuations are incorporated as random stresses in the momentum equation \cite{LandauBook} and as random fluxes in the order parameter equation \cite{SengersBook}. At equilibrium, these random fluxes are constrained by fluctuation-dissipation theorems, which relate their variances to the kinetic coefficients in the equations of motion. The fluctuation-dissipation theorem (FDT) ensures that the dynamical equations give rise to a Gibbs distribution for the fluctuating variables, as required by equilibrium statistical mechanics. Thus, together with the conservation laws, the FDT is an important constraint in the model H equations. 

The model H consists of non-linear stochastic partial differential equations which admit no analytical solutions, requiring, therefore, numerical methods of solution. Numerical methods which proceed by discretising the equations of motion on a lattice must ensure, at least, that the conservation laws and the FDTs are obeyed. This requires care as naive discretisations often violate the FDT, leaving degrees of freedom incompletely equilibrated, and therefore, without a Gibbs distribution \cite{ladd1994, metiu1983, desai1988}. 

In this paper, we solve the model H equations by combining the fluctuating lattice Boltzmann equation (FLBE) \cite{adhikari2005, dunwegladd2007} with a stochastic method of lines (SMOL) \cite{Liskovets1965, Amit2008}  using both finite difference and finite volume discretisations \cite{Capuani2004}. The formulation ensures conservation of local densities to machine precision,  and a correct balance between fluctuation and dissipation for all the degrees of freedom on the lattice. We expect our method to be widely applicable to problems in binary mixtures and other physical systems where model H is applicable, when  thermal fluctuations form an essential part of the physics \cite{stone2005, willis2009, eggers2002, Gonnella1999}. Hybrid methods have been developed in the literature in different contexts, for example in case of dynamics of binary complex fluids \cite{Gonnella2005, Orlandini2005}, but without considering thermal fluctuations. We deal with fluctuating hydrodynamics of binary fluids in detail here. Alternative schemes based on finite volume methods have also been used to simulate the  fluctuating hydrodynamics of single component fluids \cite{Donev2010} and reaction-diffusion systems \cite{Atzberger2010}. However, the methodology outlined here carries the advantages of the  lattice Boltzmann method \cite{Aidun2010} and can be generalized to other problems in fluctuating complex fluids, for instance, to the dynamics of microemulsions \cite{GompperBook} and liquid crystals.

The rest of the paper is organized as follows. In the following section we provide a detailed description of model H. We review the current understanding of solving of these equations in section \ref{sec:comparison}. In section \ref{sec:flbe} - \ref{sec:smol} we present the numerical method followed by the validation results in section \ref{sec:results}. We compare our method with previous approaches and end with a summary of  our work in section \ref{sec:summary}. 

\section{Fluctuating hydrodynamics of a binary fluid mixture}
\label{sec:theory}
We consider a coarse grained model for an isothermal binary fluid system, consisting of species $I$ and $II$ with local densities $n_I$ and $n_{II}$. The mixture as a whole has density $\rho = n_I + n_{II}$. The order parameter $\psi$, which quantifies the local composition, is taken as the normalized density difference, 
\begin{equation}
\psi = \frac{n_I - n_{II}}{n_I + n_{II}}.
\end{equation}

\subsection{Landau-Ginzburg theory}
\label{sec:energy}
The equilibrium thermodynamics of the fluid is described by the Landau free-energy functional \cite{ChaikinBook, RowlinsonBook}
\begin{equation}
F(\psi) = \int(f(\psi)+\frac{K}{2}\left|\nabla\psi|^2\right)d\mathbf{r}.
\label{eqn:Feng}
\end{equation}
Here, $\psi$  is allowed to vary beyond the limits of $\pm 1$ that follow from its definition. This ``softening'' of the order parameter has no consequence in the thermodynamic limit \cite{Wilson1974}. The first term represents the local free energy density of the bulk fluid, and is approximated as 
\begin{equation}
 f(\psi) = \frac{A}{2}\psi^2 + \frac{B}{4} \psi^4
\label{eqn:feng}
\end{equation}
with $A<0$ and $B>0$. The second term of Eq. \ref{eqn:Feng} involving the square gradient gives a free energy cost to any variation in the order parameter, and is related to the interfacial tension between the two fluid phases \cite{Kendon2001}. Minimization of Eq. \ref{eqn:feng} with respect to the order parameter gives two uniform solutions $\psi = \pm \sqrt{A/B}$, corresponding to two equilibrium fluid phases. These two phases can  coexist through a fluid interface. For a planar interface, the profile joining the two bulk phases reads
\begin{equation}
 \psi(z) = \sqrt{\frac{A}{B}}\tanh{\frac{z}{l}}
\label{eqn:tanh}
\end{equation}
where $z$ is the co-ordinate normal to the interface while
\begin{equation}
 l = \sqrt{\frac{2K}{A}}.
\label{eqn:intthick}
\end{equation}
determines the interfacial thickness. The excess energy associated to this profile with respect to the bulk energy provides the interfacial tension
\begin{equation}
 \gamma = \frac{2}{3} \sqrt{\frac{2KA^3}{B^2}}
\label{eqn:st}
\end{equation}
The corresponding chemical potential is given by the variational derivative of the free energy with respect to the order parameter  $\mu = \delta F/\delta \psi = A \psi + B \psi^3 - K \nabla^2 \psi$. The three parameters $A$, $B$, and $K$ control the interfacial thickness and interfacial energy of the mixture and after suitable non-dimensionalisations, allow for comparisons with real fluids. The additional stress due to the presence of order parameter gradients follows from the relation $\psi\nabla\mu = \nabla\cdot\boldsymbol{\sigma}^{\psi}$ \cite{anderson1998}, which is solved by
\begin{eqnarray}
 \sigma_{\alpha \beta}^{\psi} &=& K \left ( \nabla_{\alpha} \psi \right ) \left ( \nabla_{\beta} \psi \right ) \nonumber \\
&+& \delta_{\alpha \beta} \left[\psi \frac{\partial f}{\partial \psi} - f - K \psi \nabla^2 \psi - \frac{K}{2} \left |\nabla \psi \right |^2 \right].
\label{eqn:muf}
\end{eqnarray}
This additional stress includes the Laplace and Marangoni stresses due to a fluid-fluid interface. The form of this stress tensor can be motivated on the basis of an electrostatic analogy or derived directly from Poisson brackets \cite{volovick1980}.

\subsection{Model H}
\label{sec:modelH}

Model H of Halperin and Hohenberg \cite{halperin1977} describes the coupled dynamics of a conserved scalar order parameter $\psi$ and the conserved momentum density ${\bf g}$. The order parameter is described by a fluctuating Cahn-Hilliard equation, known as model B, which includes advection by fluid flow, relaxation due to chemical potential gradients, and spontaneous thermal fluctuations,
\begin{equation}
 \partial_t \psi + \nabla \cdotp \left( \mathbf{u}\psi\right) = \nabla \cdotp \left( M\nabla \mu \right) + \nabla \cdotp \hat{\boldsymbol{\xi}}.
\label{eqn:advdiff}
\end{equation}
 The mobility $M$ is the constant of proportionality in the linear phenomenological law relating the thermodynamic flux of $\psi$  to the thermodynamic force $\nabla \mu$. We consider $M$ to be a constant, though such an assumption is not necessary. Thermal fluctuations associated with $\psi$ are introduced through the random flux $\hat{\boldsymbol{\xi}}$. 

The order parameter dynamics is coupled to a fluctuating Navier-Stokes equation \cite{LandauBook} with additional stress densities arising from the order parameter. For a compressible fluid, the dynamics is governed by
\begin{eqnarray}
 \partial_t \mathbf{g} + \nabla \cdotp (\mathbf{u}\mathbf{g}) = -\nabla p + \eta \nabla^2 \mathbf{u} &+& \left [\frac{d-2}{d} \eta + \eta_b \right ] \nabla(\nabla \cdot \mathbf{u}) \nonumber \\
&+& \psi \nabla \mu + \nabla \cdotp  \hat{\boldsymbol{\sigma}} 
\label{eqn:ns}
\end{eqnarray}
together with the continuity equation for the density. In the above, $p$ stands for the isotropic contribution of the pressure, $\hat{\boldsymbol{\sigma}}$ is the random stress introduced by Landau and Lifshitz, $\psi\nabla\mu$ is the order parameter stress $\eta$ and $\eta_b$ are the shear and bulk viscosities respectively and $d$ is the dimensionality of the system.  Qualitatively, these equations describe the coupled dynamics of order parameter and flow : inhomogeneities in the order parameter generate chemical potential gradients, which in turn produce stresses in the fluid. These stresses are relaxed by fluid flow, which in turn advects the order parameter to produce inhomogeneities. 

The random flux is a zero-mean Gaussian white noise whose variance is fixed by the FDT to be
\begin{equation}
  \left \langle\hat{\xi}_{\alpha} (\mathbf{r},t) \hat{\xi_{\beta}} (\mathbf{r'},t') \right \rangle = 2 k T M \delta_{\alpha \beta} \delta(\mathbf{r}-\mathbf{r'}) \delta(t-t')
\label{eqn:fdt_orderpara}
\end{equation}
for an isothermal fluid at temperature $T$, where $k$ stands for the Boltzmann constant. Similarly, the random stress is a zero-mean Gaussian white noise whose variance is fixed by the FDT to be
\begin{equation}
 \left \langle\hat{\sigma}_{\alpha \beta}(\mathbf{r},t) \hat{\sigma}_{\gamma \delta}(\mathbf{r'},t') \right \rangle = 2 k T \eta_{\alpha \beta \gamma \delta} \delta(\mathbf{r}-\mathbf{r'}) \delta(t-t').
\label{eqn:stressfdt}
\end{equation}
where $\eta_{\alpha \beta \gamma \delta}$ is the tensor of viscosities formed out of the isotropic tensor $\delta_{\alpha\beta}$ and the shear viscosity, $\eta$, and bulk viscosity,$\eta_b$:
\begin{equation}
 \eta_{\alpha \beta \gamma \delta} = \eta\left( \delta_{\alpha \gamma} \delta_{\beta \delta} + \delta_{\alpha \delta} \delta_{\beta \gamma} \right) + \left(\eta_b - \frac{2}{d}\eta \right) \delta_{\alpha \beta} \delta_{\gamma \delta}
\end{equation}
For simplicity we assume the same viscosity for the two fluid phases.

In the next section we briefly review previous algorithms to numerically solve these coupled equations and point out why they lead to an incomplete equilibriation of both the order parameter and momentum degrees of freedom. This drawback imposes severe restrictions in the applicability of these algorithms to situations where a complete equilibriation is required, a gap which our work attempts to fill.

\section{Discretisation and FDT violation}
\label{sec:comparison}

There are ample instances in the literature where a naive discretisation of both the momentum \cite{ladd1994} and order parameter \cite{metiu1983, desai1988, sancho2000} equations have led to FDT violations on the lattice. An important question, then, is how best FDTs, derived in the continuum with respect to appropriate conservation laws, can be implemented in discrete space and time. In this section, we present a very brief survey of previous numerical schemes,  to clarify  when naive discretisations lead to  FDT violations. 

In order to gain insight into the inconsistencies associated with the order parameter discretisation, let us consider a low-order discrete representations of the divergence of a vector  $\hat{\boldsymbol{\xi}}$ and the Laplacian of a scalar $\psi$,  
\begin{eqnarray}
 \left[\nabla \cdot \hat{\boldsymbol{\xi}}\right](\mathbf{r}) &=& \sum_{i} \omega_i \mathbf{c}_i \cdot \boldsymbol{\hat{\xi}}(\mathbf{r}+\mathbf{c}_i)\\
 \left[\nabla^2 \psi\right](\mathbf{r}) &=& \sum_{i} \widehat{\omega}_i \psi(\mathbf{r}+\mathbf{c}_i).
\end{eqnarray}
Here,  $\omega_i$ and $\widehat{\omega}_i$ are weight factors which depend on the stencil, $i$ refers to the number of neighboring grid points considered, $\{\mathbf{c}_i\}$ corresponds to a lattice vector and hence $\mathbf{r}+\mathbf{c}_i$ represents the points of the chosen stencil. In a Fourier representation, they become 
\begin{eqnarray}
 \left[\nabla \cdot \hat{\boldsymbol{\xi}}\right](\mathbf{q}) &=& \sum_{i} \omega_i \mathbf{c}_i e^{i\mathbf{q} \cdot \mathbf{c}_i} \cdot \boldsymbol{\widetilde{\xi}}(\mathbf{q}) = \Gamma(\mathbf{q}) \cdot \boldsymbol{\widetilde{\xi}}(\mathbf{q})\\
 \left[\nabla^2 \psi\right](\mathbf{q}) &=& \sum_{i} \widehat{\omega}_i e^{i\mathbf{q} \cdot \mathbf{c}_i} \widetilde{\psi}(\mathbf{q}) = L (\mathbf{q}) \boldsymbol{\widetilde{\psi}}(\mathbf{q}). 
\end{eqnarray}
where $\Gamma(\mathbf{q})$ and $L(\mathbf{q})$ are the  Fourier representations of the divergence and Laplacian operators, respectively. It is easy to see that $\Gamma({\bf q})\rightarrow i{\bf q}$ and $L(\mathbf{q})\rightarrow -q^2$ as $\mathbf{q}\rightarrow 0$ for any admissible choice of stencil. In that limit, we recover the lattice analogue of the familiar relation between the gradient and Laplacian operators, so that $L(\mathbf{q}) = \Gamma(\mathbf{q}) \cdot \Gamma(\mathbf{q})$. At high wavenumbers, however, this relation is no longer true. Indeed, it is violated by all standard nearest neighbour stencils \cite{AbramowitzBook}.

To see how this affects discretisations of the fluctuating Cahn-Hilliard equation, we linearize Eq. \ref{eqn:advdiff} about a state of zero flow, for completely local and harmonic free energy ($B=0, K=0$ in Eq. \ref{eqn:Feng}), with a mobility that is independent of the order parameter.  Discretising and Fourier transforming, we obtain
\begin{equation}
 \partial_t \widetilde{\psi}(\mathbf{q}) = M L(\mathbf{q}) A \widetilde{\psi}(\mathbf{q}) + \Gamma(\mathbf{q}) \cdot \widetilde{\xi}(\mathbf{q}).
\label{eqn:FTorderpara}
\end{equation}
It is evident from Eq. \ref{eqn:FTorderpara} that fluctuations in the order parameter equation will satisfy  the FDT of Eq. \ref{eqn:fdt_orderpara} on the lattice \textit{if and only if} $L(\mathbf{q}) = \Gamma(\mathbf{q}) \cdot \Gamma(\mathbf{q})$. Equivalently, the discrete operators should satisfy $\nabla^2 = \nabla. \nabla$ in real space. Since this is not true for the standard choices of the previous operators \cite{AbramowitzBook}, resulting discretisations violate FDT.

\begin{figure}
\includegraphics[width=1.0\linewidth]{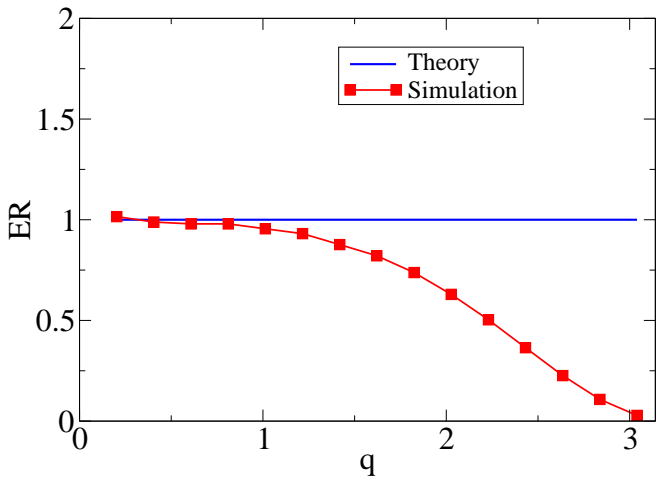}
\caption{(Color Online) Equilibrium ratio (ER) according to Eq. \ref{eqn:test1} as a function of wave vector magnitude $q$ along the diagonal $x=y=z$ from using a conventional method \cite{metiu1983} based on finite difference discretisation for both the divergence and Laplacian operators. Simulation results show significant differences with theoretical predictions at large wavenumbers.}
\label{fig:test1wrong}
\end{figure}

To verify the above analysis, we perform simulations using the method proposed by Petschek and Metiu \cite{metiu1983} and used, for example, in \cite{desai1988} and \cite{sancho2000}. Their method is essentially the one outlined above, with specific choices of the gradient and Laplacian. Simulations are carried out on a $32 \times 32 \times 32$ domain with a cubic grid and unit spacing, $\Delta x = 1$ and unit time step $\Delta t = 1$ using a stochastic Runge-Kutta algorithm \cite{wilkie2004}.  We compare the theoretical value of the Fourier mode amplitudes of order parameter as given by the Gibbs distribution
\begin{equation}
 \langle|\widetilde{\psi}(\mathbf{q})|^2 \rangle = \frac{kT}{A}
\label{eqn:test1}
\end{equation}
with our simulation data. We define the equilibrium ratio ER as the ratio of simulated values to the theoretical value. If all Fourier modes are in equilibrium, the ER will be unity as dictated by Eq. \ref{eqn:test1}. The results obtained are displayed in Fig. \ref{fig:test1wrong}. As can be seen, the difference from the expected theoretical value of ER $= 1$ is quite significant : the match is restricted to only small wave numbers and clearly shows the breakdown of FDT at high wavenumbers.
Having identified the spatial discretisation as the main source of error in FDT violation on the lattice, we will analyze in the next section how to circumvent it for a scalar order parameter. Its generalization  to vector and tensor order parameters is straightforward. 

Fluctuations have also been included in lattice Boltzmann equation (LBE) to recover fluctuating Navier-Stokes equations. Ladd \cite{ladd1994} proposed a modification of the LBE with the addition of fluctuating stresses. A Langevin interpretation of the Boltzmann equation then yields the equations of fluctuating hydrodynamics \cite{LandauBook}. However, as was pointed out in \cite{adhikari2005}, Ladd's method ensures thermalisation only in the small wave number limit. This was resolved by relating thermal fluctuations to all sources of dissipation associated with the collision operator in the lattice Boltzmann equation, leading to thermal equilibrium for all modes,  including the ghost modes \cite{SucciBook}.  This was confirmed subsequently in \cite{dunwegladd2007}. The fluctuating lattice Boltzmann equation (FLBE) \cite{adhikari2005, dunwegladd2007} provides a consistent lattice discretisation for the Navier-Stokes equations and is the approach we shall use in this work. 

The FLBE approach has recently been generalized to hydrodynamic fluctuations of non-ideal gases \cite{Gross2010,Grossarxiv}, but only a few studies have addressed  thermal fluctuations in binary mixtures in the context of LBE. Noise driven spinodal decomposition was studied in \cite{Gonnella1999} by combining Ladd's fluctuating LBE with a fluctuating kinetic equation for the order parameter. However, this method 
does not respect FDT for either the momentum or the order parameter. Extending this to binary fluids maintaining FDT is considerably more difficult and so we prefer the alternative hybrid method described below. 

\section{Fluctuating Navier-Stokes solver}
\label{sec:flbe}
We use the FLBE method for solving the fluctuating Navier-Stokes equations. The FLBE  introduced in \cite{adhikari2005} needs to be modified to include force densities, which in the hybrid method, are the divergences of order parameter stresses. Since the combination of noise and external force densities \cite{nash2008} modifies the moment relations between the distribution functions and the hydrodynamic variables, we discuss now  the main new features of FLBE and provide a detailed and self-contained derivation in Appendix \ref{sec:flbeappen}.

In a standard $DdQn$ LBE model where the velocity space is discretized into $n$ components in $d$ dimensional space, the discrete form of the fluctuating Boltzmann equation reads \cite{Grossarxiv}
\begin{equation}
 \partial_t f_i + \mathbf{c}_i \cdot \nabla f_i + [{\bf F}\cdot\nabla_{\bf c}f]_i= - \sum_j L_{ij}( f_j -f_j^0)+ \zeta_i
\label{eqn:flbe}
\end{equation}
where $\mathbf{F}(\mathbf{x},t)$ is an effective force density, $\zeta_i(\mathbf{x},t)$ stands for the fluctuations in the populations, and $L_{ij}$ is the discrete form of the collision integral and is related to the fluid viscosity. The moments of the single particle distribution function $f_i$, defined at lattice node $\bf{x}$ with velocity $\mathbf{c}_i$ at time $t$, give the fluid mass, momentum and stress densities:
\begin{equation}
 \rho = \sum_{i=0}^{n} f_i, \hspace{2mm} \rho\mathbf{v} = \sum_{i=0}^{n} f_i\mathbf{c}_i, \hspace{2mm}S_{\alpha \beta} = \sum_{i=0}^{n}{f}_i Q_{i \alpha \beta}
\label{eqn:momreln}
\end{equation}
where $Q_{i\alpha\beta} = c_{i\alpha}c_{i\beta} - c_s^2 \delta_{\alpha\beta}$. The collision operator $L_{ij}$ controls the relaxation of $f_j$ to  equilibrium,  $f_j^0$. We can take advantage of the hyperbolic character of the FLBE and use the method of characteristics to evolve Eq. \ref{eqn:flbe} over a finite time step. When accounting for the effect of forces and fluctuations in the evolution of $f_i$, it is convenient to introduce the auxiliary distribution function
\begin{equation}
 \bar{f}_i(\mathbf{x},t) = f_i(\mathbf{x},t) - \frac{\Delta t}{2} R_i(\mathbf{x},t)
\label{eqn:barreln}
\end{equation}
in terms of  $R_i(\mathbf{x},t) = - \sum_j L_{ij}( f_j -f_j^0)+ \Phi_i$, which represents the effects of collision, forcing and thermal fluctuations, see Appendix \ref{sec:flbeappen}. For a single-time relaxation operator, $L_{ij} = \delta_{ij}/\tau$, the hydrodynamic variables are related to the auxiliary distributions as
\begin{eqnarray}
 \rho &=& \sum_{i=0}^{n} \bar{f}_i\\
\rho v_{\alpha} &=& \sum_{i=0}^{n} \bar{f}_i c_{i\alpha} + \rho F_{\alpha} \frac{\Delta t}{2}\\
S_{\alpha \beta} &=& \sum_{i=0}^{n}\bar{f}_i Q_{i \alpha \beta} + \frac{\Delta t/2}{\tau + \Delta t/2} \left(-\sum_{i=0}^{n}\bar{f}_i Q_{i \alpha \beta} \right. \nonumber \\
&+& \left. \rho v_{\alpha} v_{\beta} + \tau(v_{\alpha}F_{\beta}+F_{\alpha} v_{\beta}) + \tau \sum_{i=0}^{n}\zeta_i Q_{i\alpha\beta}\right).
\label{eqn:stress}
\end{eqnarray}
where the equilibrium distribution, $f_i^0$, can be reconstructed from $\rho$ and $\rho \mathbf{v}$. In Eq. (\ref{eqn:stress}) $\sum_{i=0}^{n}\zeta_i Q_{i\alpha\beta} $ is the fluctuating contribution to the stress. 

The effective force density is the divergence of the order parameter stress  
\begin{equation}
\mathbf{F} = \nabla \cdot \boldsymbol{\sigma}^{\psi} = \psi \nabla \mu
\end{equation}
which can be verified using Eq. \ref{eqn:muf}. To compute this force density  we use a symmetrized, second order accurate nearest-neighbor central difference stencil for the gradient
\begin{eqnarray}
 \nabla \mu(x,y,z) &=& \frac{1}{2} \left[ \mu(x+1,y,z) - \mu(x-1,y,z) \right] \hat{x}\nonumber\\
&+& \frac{1}{2} \left[ \mu(x,y+1,z) - \mu(x,y-1,z) \right]\hat{y} \nonumber\\
&+& \frac{1}{2} \left[ \mu(x,y,z+1) - \mu(x,y,z-1) \right]\hat{z}
\end{eqnarray}
and the Shinozaki-Oono discretisation \cite{Shinozaki1993} of the Laplacian (Eq. \ref{eqn:solap}) to calculate $\nabla^2\psi$ in the chemical potential.

\section{Fluctuating Cahn-Hilliard solver}
\label{sec:smol}
We use a stochastic method of lines (SMOL) discretisation \cite{Amit2008} to solve the fluctuating Cahn-Hilliard equation for the order parameter. Since it does not contain a pressure term which acts as a Lagrange multiplier in the incompressible Navier-Stokes equations, there is no particular benefit in using a kinetic algorithm with its large number of degrees of freedom in solving for a single scalar variable. Here, we adopt a semi-discretisation strategy \cite{Liskovets1965, Amit2008}, discretising the spatial variables to obtain a set of coupled stochastic ordinary differential equations. The spatial discretisations we propose ensure that the conservation law is respected to machine precision and that the fluctuation and dissipation are in balance for all wave vectors. We propose a finite-difference and finite-volume discretisations, discussing their relative merits below. The temporal integration of the resulting stochastic differential equations is done using a Runge-Kutta algorithm proposed recently by Wilkie \cite{wilkie2004}. This is a straightforward generalization of the deterministic Runge-Kutta algorithm where the noise is held constant through the integration step. The methodology may be improved using implicit schemes to increase the accuracy.

\subsection{Finite difference method}
\label{sec:FD}
To proceed towards a discretisation of the  fluctuating Cahn-Hilliard equation which preserves FDT, we write the order parameter evolution equation in Fourier space
\begin{equation}
 \partial_t \widetilde{\psi}(\mathbf{q}) =   M L(\mathbf{q})\widetilde{\mu}(\mathbf{q}) + 
\Gamma(\mathbf{q}) \cdot \widetilde{\xi}(\mathbf{q})
\end{equation}
assuming a constant mobility. Defining the divergence of the noise in Fourier space as $\widetilde{\eta}(\mathbf{q}) = \Gamma(\mathbf{q}) \cdot \widetilde{\xi}(\mathbf{q})$, we see that it must satisfy 
\begin{equation}
\left \langle\hat{\eta} (\mathbf{q}, t) \hat{\eta} ({\mathbf{q}'},t') \right \rangle = - 2 k T M L(\mathbf{q}) \delta(\mathbf{q}+\mathbf{q'}) \delta(t-t').
\label{eqn:etafdt}
\end{equation}
Instead of constructing a divergence operator $\Gamma(\mathbf{q})$ which satisfies  $\Gamma(\mathbf{q})\cdot\Gamma(\mathbf{q}) = L(\mathbf{q})$ we directly use the above relationship to construct the noise in Fourier space. This is then inverse-transformed to real space to provide a noise which has correlations compatible with the discretisation of the Laplacian and the same Laplacian stencil is used to calculate $\nabla^2\mu$ and $\nabla^2\psi$. The generation of noise in Fourier space has been used earlier in spectral methods \cite{Atzberger2007} to respect FDT in discrete space.

It is important to ensure as isotropic a discretisation of the Laplacian as possible, to avoid artifacts like spurious pinning of interfaces by the lattice. We have compared in Appendix \ref{sec:lap} four standard finite-difference stencils reported in the literature, see Fig. \ref{fig:laps} in Appendix \ref{sec:lap} where expressions for their Fourier transforms $L(\mathbf{q})$ are also provided. The Laplacian of Shinozaki and Oono \cite{Shinozaki1993} is the most isotropic one and we use it for our discretisation. The advective flux, $\nabla \cdotp (\mathbf{u}\psi)$, is discretized using a second order accurate, conservative, central difference scheme
\begin{eqnarray}
 &\left[ \nabla \cdot (\mathbf{u}\psi)\right]&(x,y,z) \nonumber\\
&=& \frac{1}{2} \left\{ [u_x \psi](x+1,y,z) - [u_x \psi](x-1,y,z) \right\} \nonumber\\ 
 &+& \frac{1}{2} \left\{ [u_y \psi](x,y+1,z) - [u_y \psi](x,y-1,z) \right\}  \nonumber\\
 &+& \frac{1}{2} \left\{ [u_z \psi](x,y,z+1) - [u_z \psi](x,y,z-1) \right\}.
\end{eqnarray}

\subsection{Finite volume method}
\label{sec:FV}

\begin{figure}
\includegraphics[width=0.5\linewidth]{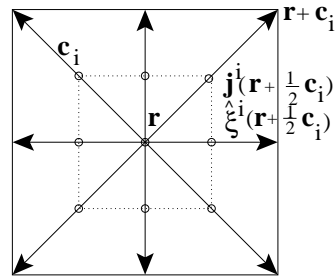}
\caption{Illustration of the stencil used for the numerical tests in the finite volume method for a two dimensional case. This stencil corresponds to the  D2Q9 lattice Boltzmann model. Physical quantities ,e.g. $\psi, \mu, \nabla \mu,$ and $\mathbf{u}$, are defined at node $\mathbf{r}$ which has its neighbors at $\mathbf{r} + \mathbf{c}_i$. All fluxes $\mathbf{j^i},\boldsymbol{\xi}^i$ (diffusive, convective and random) are defined at the mid point of the links ($\mathbf{r} + \frac{1}{2} \mathbf{c}_i$) connecting $\mathbf{r}$ and $\mathbf{r} + \mathbf{c}_i$. (See Eq. \ref{eqn:fvdiff} - \ref{eqn:fvgrad})}
\label{fig:fvcube}
\end{figure}

It is possible to formulate an alternative discretisation for the fluctuating Cahn-Hilliard equation, based on a finite-volume formulation. Such an approach, using fluxes defined on lattice links, has been proposed to study the electrokinetic equations in the absence of fluctuations in \cite{RotenBerg2008, RotenBerg2010}. Alternative finite volume schemes may also be found in the context of reaction-diffusion systems \cite{Atzberger2010}. Specifically, we choose a $DdQn$ cubic lattice and a set of link vectors $\{\mathbf{c}_i\}$ as done usually with lattice Boltzmann models. Thus, for any node $\mathbf{r}$, the set of points $\mathbf{r}+\mathbf{c}_i$ are also lattice nodes. The divergence at a node $\mathbf{r}$ is then written as a sum of fluxes $\mathbf{j}^i$ defined on the midpoint $\mathbf{r} + \frac{1}{2}\mathbf{c}_i$ of the link connecting the node to its neighbour $\mathbf{r}+\mathbf{c}_i$. This is schematically represented in Fig. \ref{fig:fvcube} for $D2Q9$. . Then, Eq. \ref{eqn:advdiff} can be discretised as
\begin{equation}
 \partial_t \psi = \sum_i w_i \mathbf{c}_i\cdot \mathbf{j}^i + \sum_i w_i \mathbf{c}_i\cdot \hat{\boldsymbol{\xi}^i}
\label{eqn:fvdiff}
\end{equation}
where $w_i$ are a normalised set of weights that ensure isotropy and $\mathbf{j}^i$ and $\hat{\boldsymbol{\xi}^i}$ are the deterministic and random contributions to the order parameter flux, respectively. This ensures the conservation of the order parameter to machine accuracy. 

The choice of expressions which relate the fluxes to the densities at the nodes must now be dictated by the requirement that the FDT holds for all wavevectors. We use a symmetric average of node values to compute the mid-point fluxes \cite{Capuani2004},
\begin{eqnarray}
 \mathbf{j}^{i} &=& M \frac{1}{2} \left[\nabla \mu(\mathbf{r}) + \nabla \mu( \mathbf{r}+\mathbf{c}_i) \right] \nonumber\\
&&- \frac{1}{2} \left[({\mathbf u}\psi)(\mathbf{r}) + (\mathbf{u}\psi)( \mathbf{r}+\mathbf{c}_i) \right]
\label{eqn:fvflux} \\
\hat{\boldsymbol{\xi}^i} &=& \frac{1}{2} \left[\hat{\boldsymbol{\xi}}(\mathbf{r}) + \hat{\boldsymbol{\xi}}(\mathbf{r}+\mathbf{c}_i) \right].
\label{eqn:noiseflux}
\end{eqnarray}
Here, $c_s^2\delta_{\alpha\beta} = \sum_i c_{i\alpha}c_{i\beta}$. To be consistent with this choice, the gradient of the chemical potential must be computed using 
\begin{equation}
 \nabla \mu (\mathbf{r}) = \frac{1}{c_s^2} \sum_i w_i \mathbf{c}_i \mu(\mathbf{r} +\mathbf{c}_i).
\label{eqn:fvgrad}
\end{equation}
It is only with the combined choice of the divergence, symmetric averaging, and the gradient that the fluctuating Cahn-Hilliard equation takes the form
\begin{equation}
 \partial_t \psi(\mathbf{q}) + \mathbf{\Gamma}(\mathbf{q})\cdot(\mathbf{u}\psi)(\mathbf{q}) = \mathbf{\Gamma}(\mathbf{q}) \cdot [M \mathbf{\Gamma}(\mathbf{q}) \mu(\mathbf{q})) + \boldsymbol{\xi}(\bf{q})]
\label{eqn:fvft}
\end{equation}
where $\mathbf{\Gamma}(\mathbf{q}) \equiv \sum_i w_i \mathbf{c}_i \exp(i \mathbf{q} \cdot \mathbf{c}_i) $ is the representation of the $\nabla$ operator on the lattice. Our choice of discretisation ensures that the same operator  $\mathbf{\Gamma}(\mathbf{q})$ appears in both the gradient and the divergence in the diffusive term in the Cahn-Hilliard equation. As a result, $\nabla \cdot \nabla = \nabla^2$ is preserved at all wavevectors, and not only when $\mathbf{q}\rightarrow 0$ as happens with standard discretisations. The resulting Laplacian $[L(\mathbf{q})]_{FV} = \mathbf{\Gamma}(\mathbf{q})\cdot\mathbf{\Gamma}(\mathbf{q})$ is less isotropic than the Shinozaki-Oono Laplacian as shown in Fig. \ref{fig:lapsfv} in Appendix \ref{sec:lap}. Therefore, we use the Shinozaki-Oono Laplacian to calculate $\nabla^2\psi$ in the chemical potential.

Compared to the finite-difference method of the previous section, the finite-volume method is not restricted to periodic geometries, and thus allows for simulations with wall or shear boundary conditions. The computational overhead is significantly reduced since the expensive Fourier construction of the noise is no longer required.

\section{Results and Validation}
\label{sec:results}

The order parameter  induces a force on the fluid, accelerating it while the fluid, in turn, advects the order parameter. Although this requires, in principle, an algorithm which updates self-consistently  both fields, we have to do it sequentially at every time step due to the coupling of two different methods, resulting in a hybrid scheme for the model H equations. However, we have not found any event where  the proposed algorithm of alternate marching in time of FLBE and SMOL  leads to spurious cross correlations between momentum and order parameter fluctuations.

A number of tests have been carried out to validate the algorithm including static and dynamic correlations for the order parameter and standard tests for hydrodynamics. We have always used a D3Q15 model for FLBE  with lattice units $\Delta x = \Delta t = 1$ which leads to a speed of sound $c_s = \sqrt{1/3}$. To ensure that compressibility is negligible, we work in parameter regimes where the Mach number is small, $Ma = u/c_s \ll 1$.  Except when otherwise stated, all simulations have been performed on a $32 \times 32 \times 32$ lattice which  is initialized with a uniform random distribution, and statistics are collected once the system has equilibrated. The relaxation parameter, $\tau=1.1$ and the temperature $kT=1/3000$ are used \cite{adhikari2005} in FLBE unless otherwise specified. Periodic boundary conditions are used in all directions in all the simulations.

\subsection{Order parameter fluctuations}

We analyze initially a miscible mixture without surface tension, characterized by $B=K=0$. Since in this case the free energy functional, Eq. \ref{eqn:Feng} ,  is parabolic, the equilibrium order parameter distribution follows the Gibbs distribution with Gaussian order parameter fluctuations of amplitude given in Eq. \ref{eqn:test1}.


\begin{figure}
\includegraphics[trim = 0mm 0mm 0mm 0mm, clip, width=1.0\linewidth]{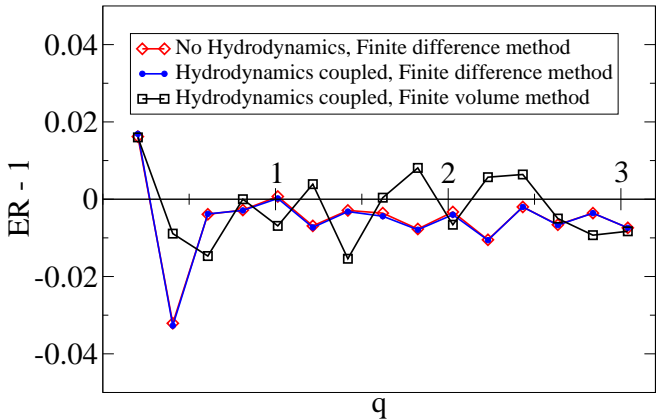}
\caption{(Color Online) Error in the equilibrium ratio as a function of wave vector magnitude, $q$, along the diagonal $x = y = z$ considering (1) diffusion alone and (2) coupled hydrodynamics with (i) finite difference and (ii) finite volume method for the quadratic free energy functional (see Eq. \ref{eqn:test1}). Simulations have been done on a $32 \times 32 \times 32$ lattice with equilibrium initial conditions and parameters used are $A=0.625$, $B=0$, $K=0.0$ and $M=0.095$. Ensemble averaging is done over $10^4$ time steps and over 25 realizations.}
\label{fig:test1a}
\end{figure}

\begin{figure}
\includegraphics[trim = 0mm 0mm 0mm 0mm, clip, width=1.0\linewidth]{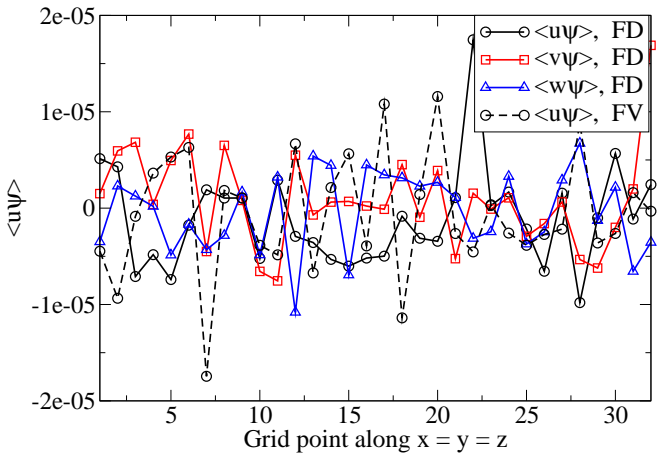}
\caption{(Color Online) Velocity-order parameter correlation for all three components of the velocity in Cartesian coordinates  along the diagonal $x = y = z$ of the domain considering coupled hydrodynamics and using the  finite difference scheme  for the quadratic free energy functional (see Eq. \ref{eqn:test1}) and for the same set of parameters as in Fig. \ref{fig:test1a}. Ensemble averaging is done over $10^4$ time steps and over 25 realizations. No cross correlations are present between fields of different tensorial nature. The results obtained using finite volume method are shown only for  one velocity component  for clarity.}
\label{fig:test1upsi}
\end{figure}

\begin{figure*}
\subfigure[]{\includegraphics[trim = 20mm 10mm 20mm 0mm, clip, width=0.32\linewidth]{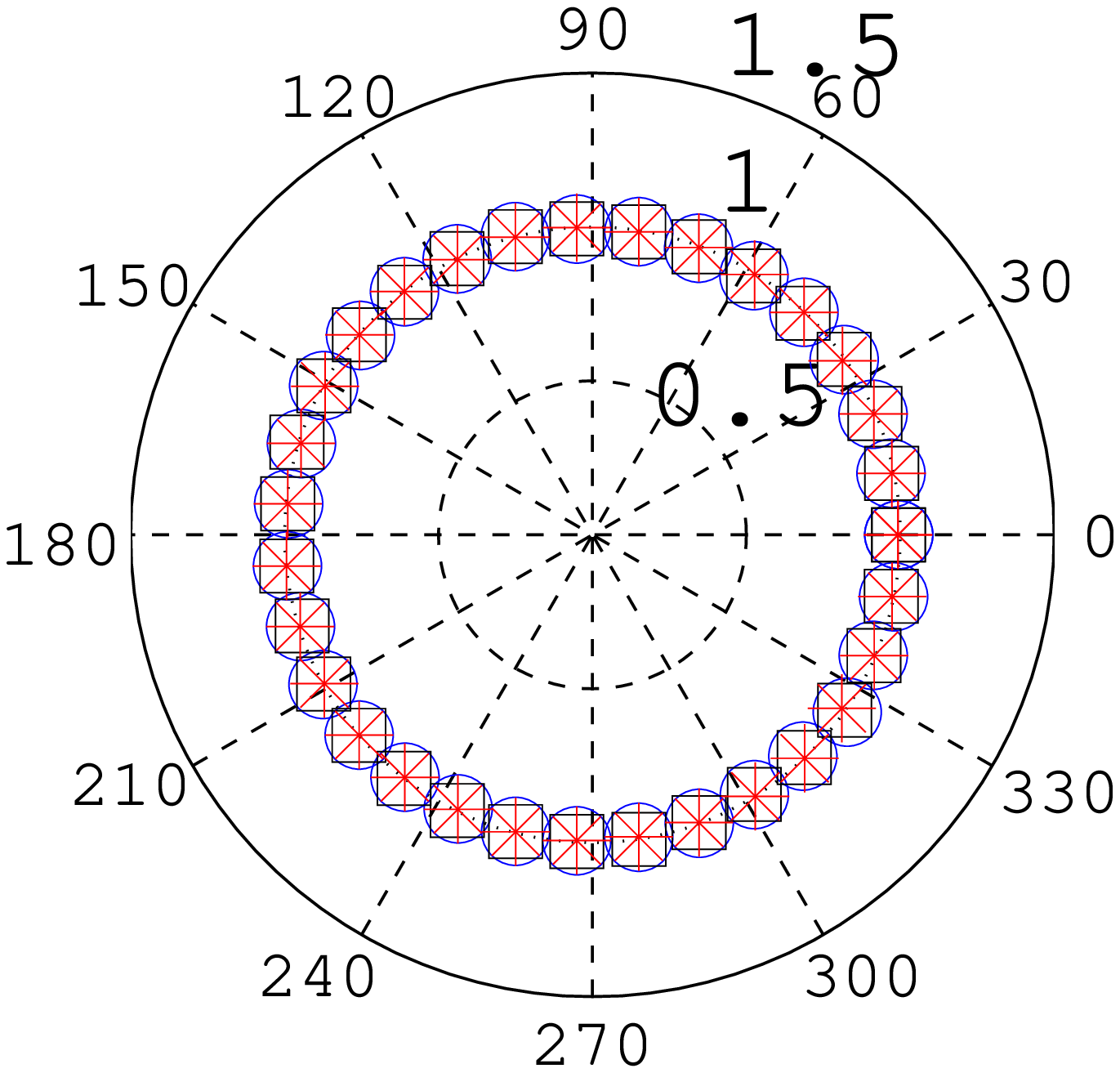}}
\subfigure[]{\includegraphics[trim = 20mm 10mm 20mm 0mm, clip, width=0.31\linewidth]{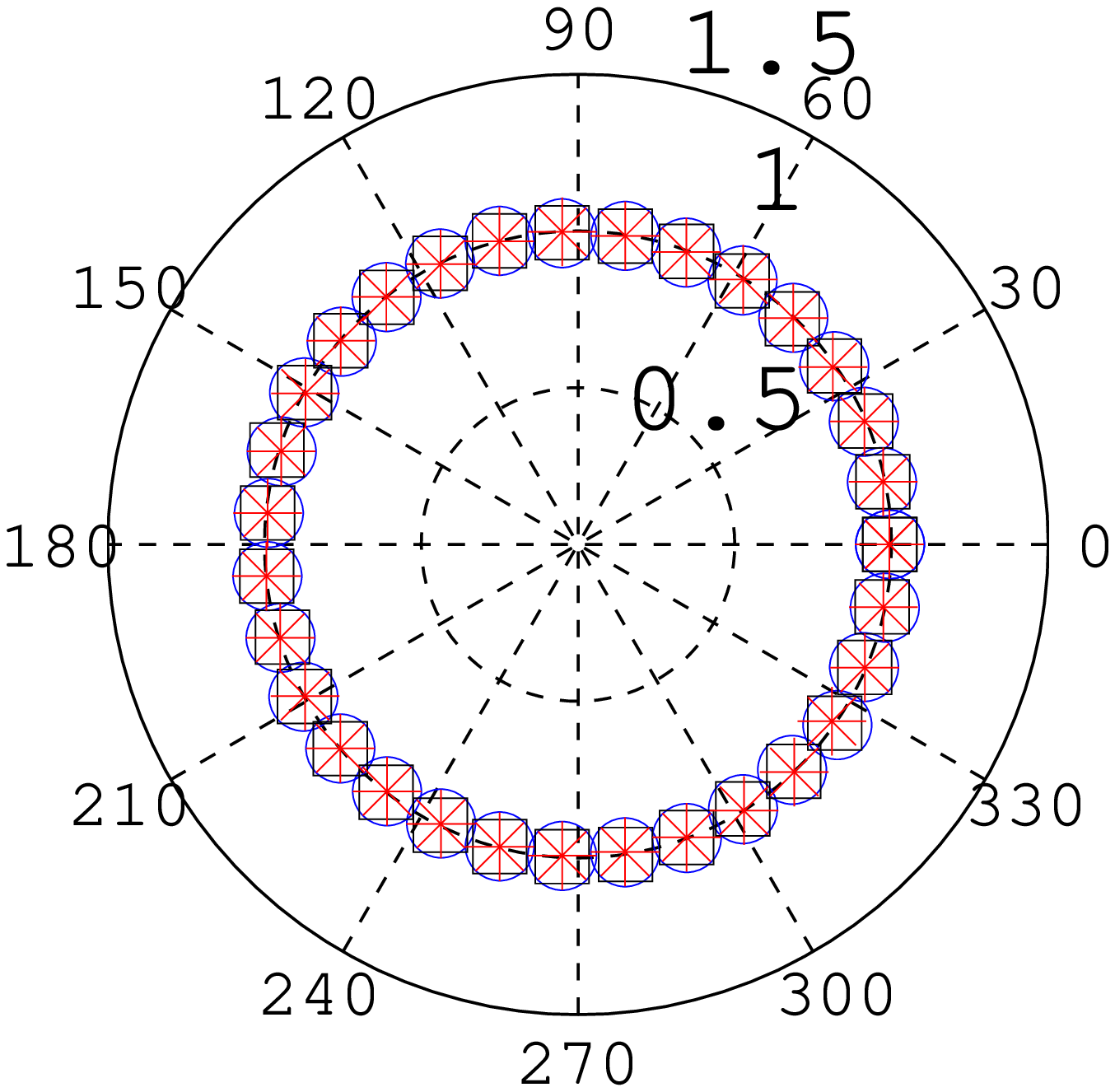}}
\subfigure[]{\includegraphics[trim = 20mm 10mm 20mm 0mm, clip, width=0.31\linewidth]{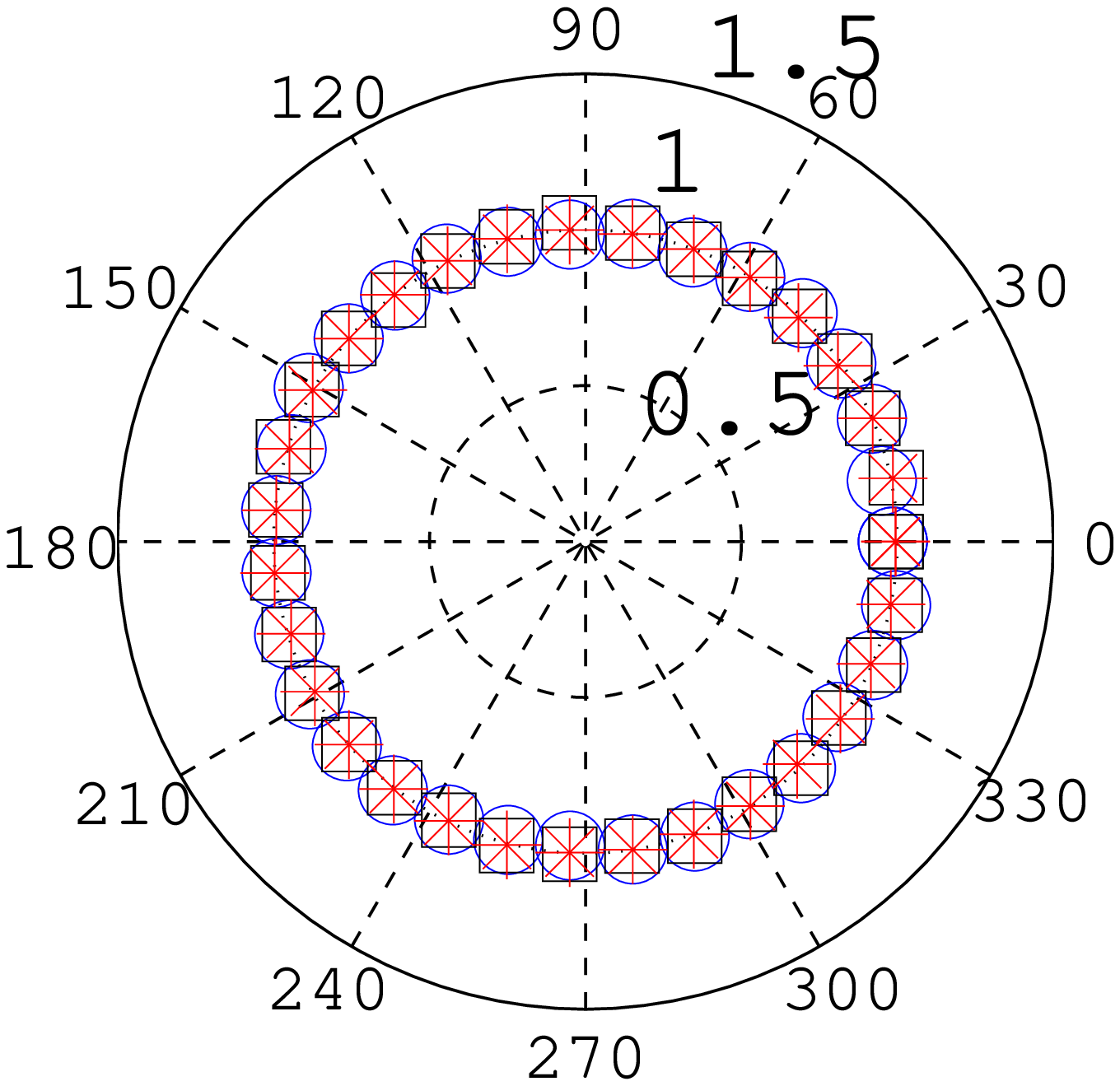}}
\caption{(Color Online) Polar plots where the radius indicates the ER as a function of azimuthal angle on lattice points of a fixed modulus (16 lattice units from the center), i.e. along a ($\cos{\theta}, \sin{\theta}, z=\text{constant}$) for (a) diffusion alone, finite difference method and (b) hydrodynamics coupled with finite difference method, (c) hydrodynamics coupled with finite volume method. Three different symbols \textcolor{blue}{o}, \tiny $\square$, \normalsize{ \textcolor{red}{\textasteriskcentered}}~correspond to $z=N/8, 2N/8$ and $3N/8$ planes respectively. Data obtained from the same simulations used in Fig. \ref{fig:test1a}.}
\label{fig:test1b}
\end{figure*}

Fig. \ref{fig:test1a} displays the error in the equilibrium ratio (ER) between the measured static correlation functions of the order parameter and the theoretical prediction, Eq. \ref{eqn:test1} independent of the wave vector magnitude, as a function of the magnitude of the wave vector, for $q_x=q_y=q_z$, both without and with hydrodynamic coupling. In the latter situation we have also compared the performance of the finite difference method (section \ref{sec:FD}) and the finite volume method (section \ref{sec:FV}). In all cases we obtain an excellent agreement for the entire wave vector spectrum, as opposed to the spurious deviations observed in Fig. \ref{fig:test1wrong} for a standard discretisation of Eq. \ref{eqn:advdiff}. In Fig. \ref{fig:test1upsi} the velocity-order parameter correlations are plotted using both the finite difference and finite volume method to show that no spurious scalar-tensor correlations  develop in the proposed numerical scheme.

In order to check the homogeneity and isotropy of the fluctuations, polar plots  are shown in Fig. \ref{fig:test1b}. In these plots, the radius represents the ER as a function of the azimuthal angle in a given $z$ plane in lattice space. Different symbols correspond to three different $z$ planes. ER remains essentially unity in all cases, indicating that FDT is satisfied in all directions in the lattice.

\begin{figure}
\subfigure[]{\includegraphics[trim = 15mm 0mm 20mm 0mm, clip, width=1.0\linewidth]{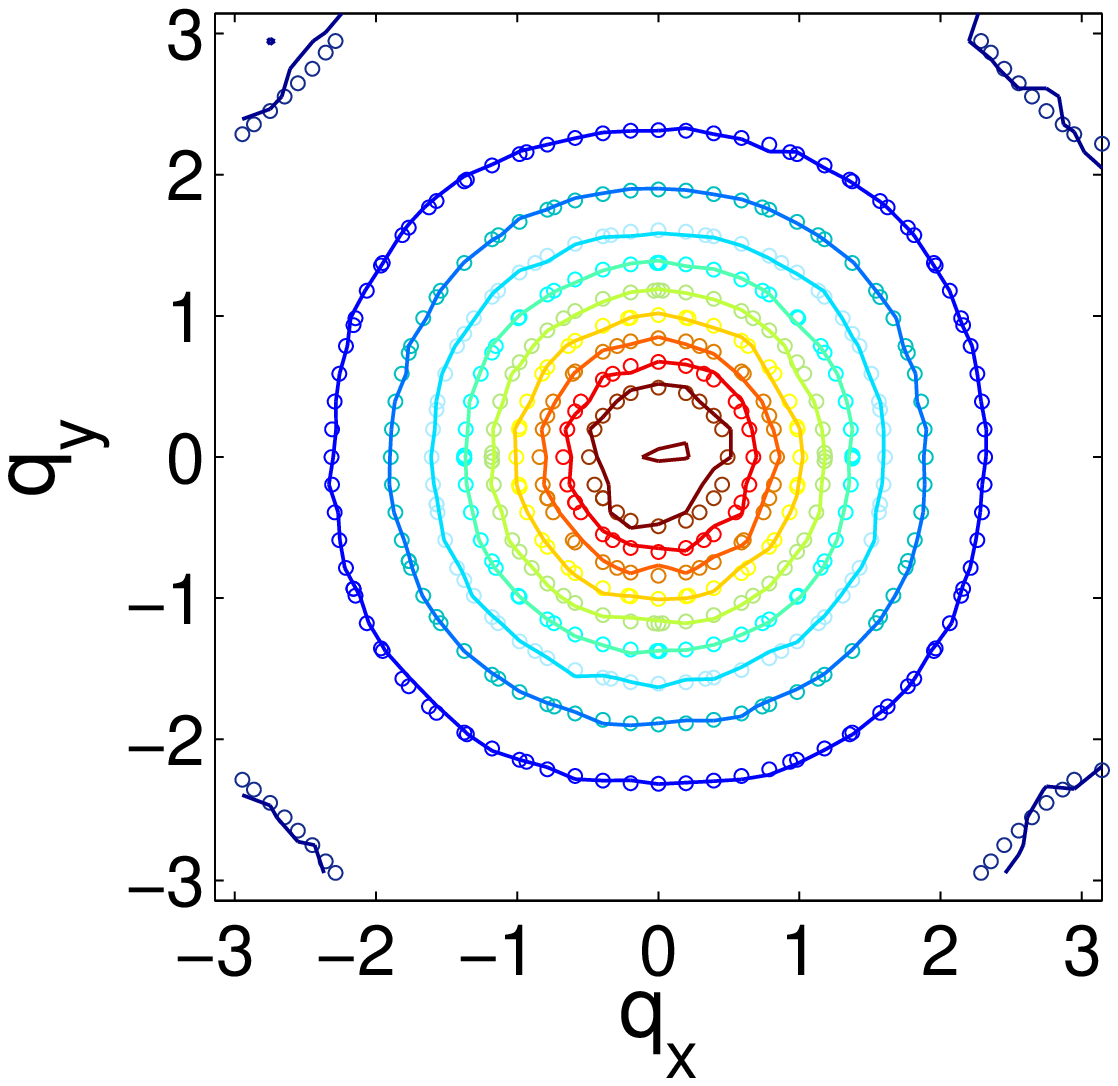}}
\caption{(Color Online) Constant $\langle|\widetilde{\psi}(q)|^2 \rangle$ values obtained at equilibrium, from a simulation considering diffusion alone (without any coupling to hydrodynamics) for the free energy functional described by Eq. \ref{eqn:Feng} with $B=0$. Simulations obtained using the finite difference method. Results are shown in a wave number plane of ($q_x,q_y$). Analytical expression from Eq. \ref{eqn:test2} are superposed onto it using symbols for comparison.  Simulations are performed on a $32 \times 32 \times 32$ lattice with equilibrium initial conditions, $A=0.025$, $K=0.01$ and $M=0.1$. Ensemble averaging is done over $10^5$ time steps and over 25 realizations.}
\label{fig:test2a}
\end{figure}

\begin{figure*}
\subfigure[]{\includegraphics[trim = 15mm 0mm 20mm 0mm, clip, width=0.48\linewidth]{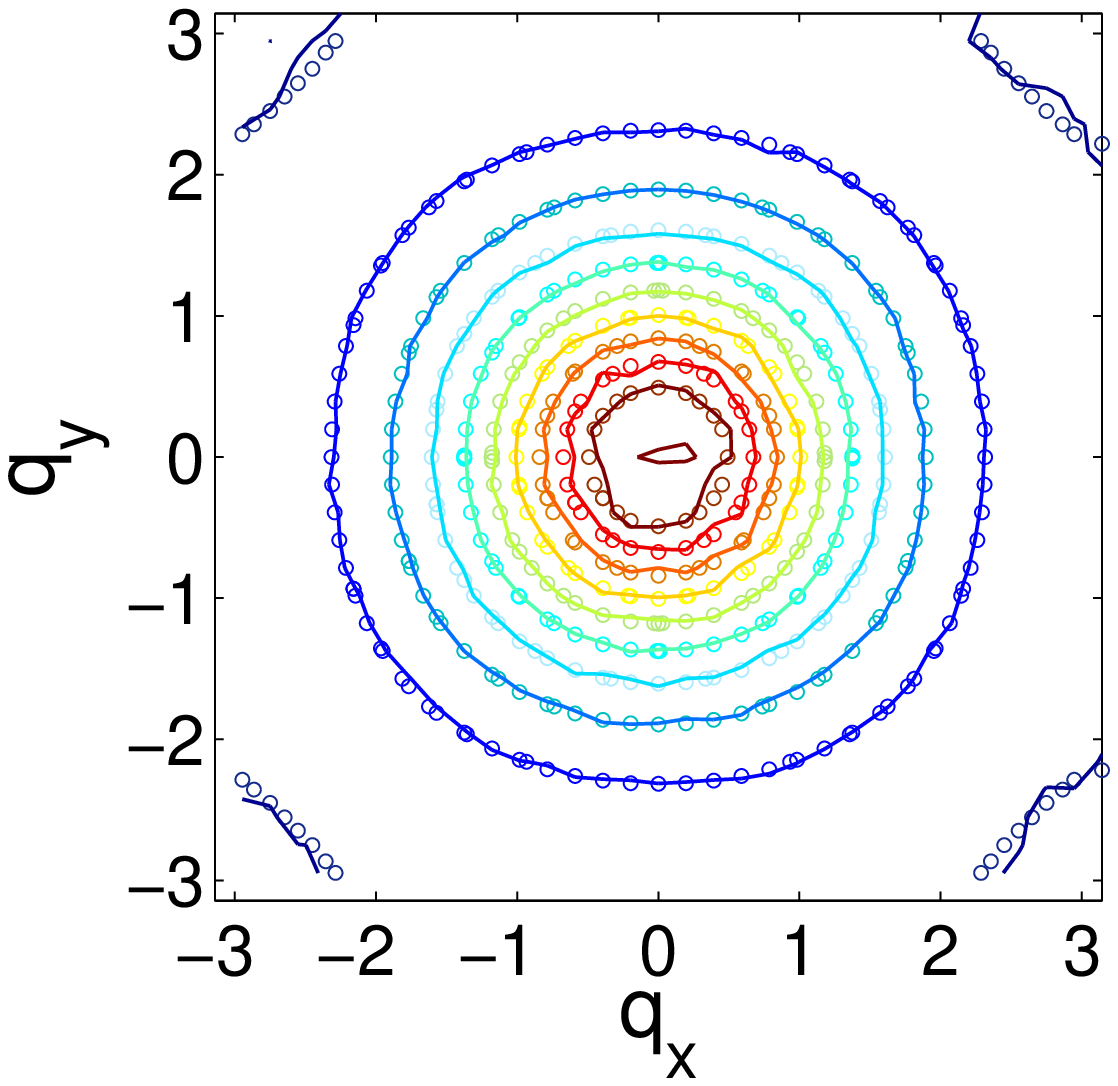}}
\subfigure[]{\includegraphics[trim = 15mm 0mm 20mm 0mm, clip, width=0.48\linewidth]{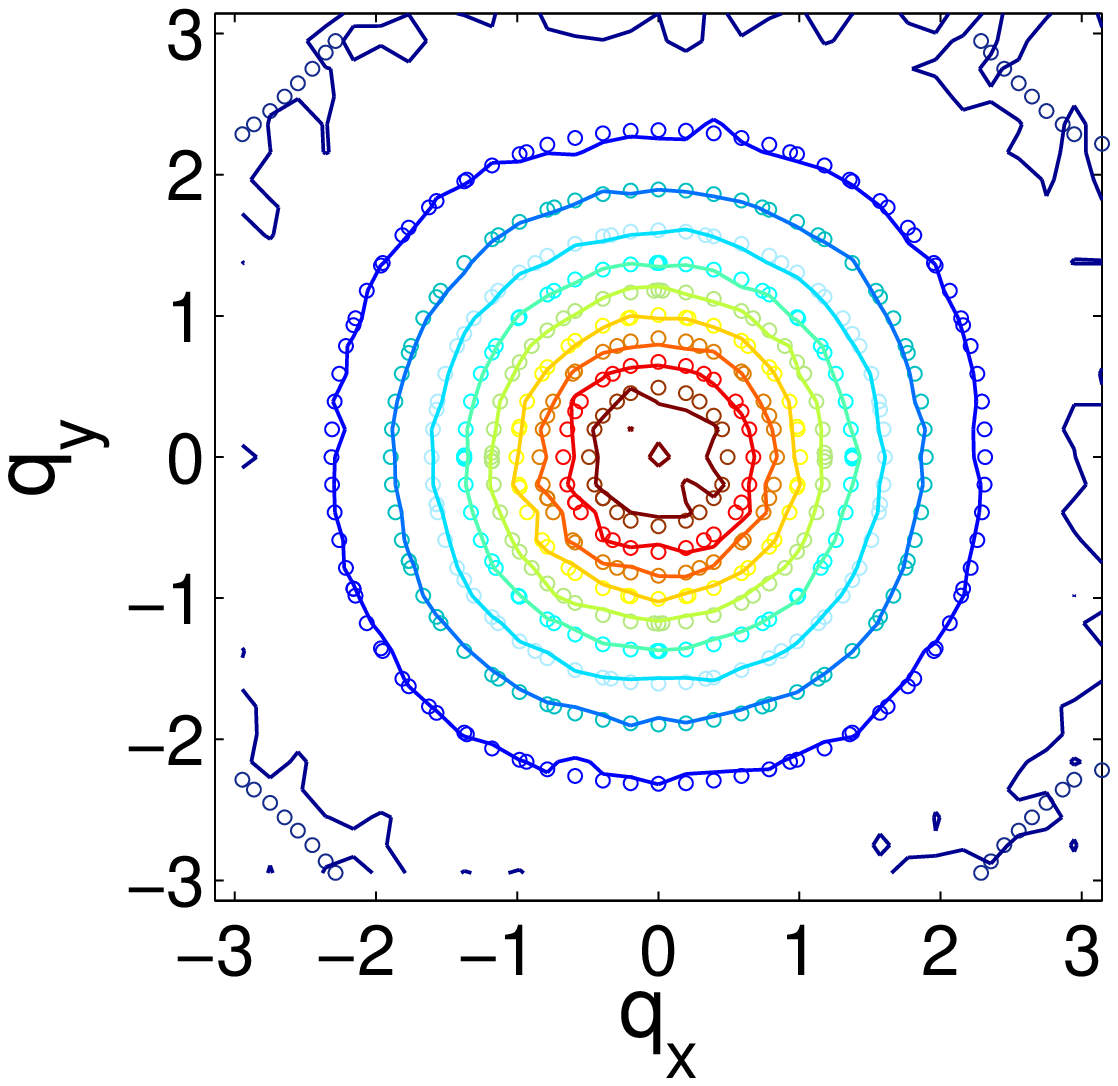}}
\caption{(Color Online) Constant values of $\langle|\widetilde{\psi}(q)|^2 \rangle$ from the simulations when the dynamics of the order parameter is coupled to the fluid dynamics for the same parameters and lattice size used in Fig. \ref{fig:test2a}. Results for both the finite difference method (a)  and the finite volume method (b) are shown at a constant $q_z$ plane and expected values from Eq. \ref{eqn:test2} are superposed as symbols.}
\label{fig:test2}
\end{figure*}

\begin{figure}
\includegraphics[trim = 0mm 0mm 0mm 0mm, clip, width=1.0\linewidth]{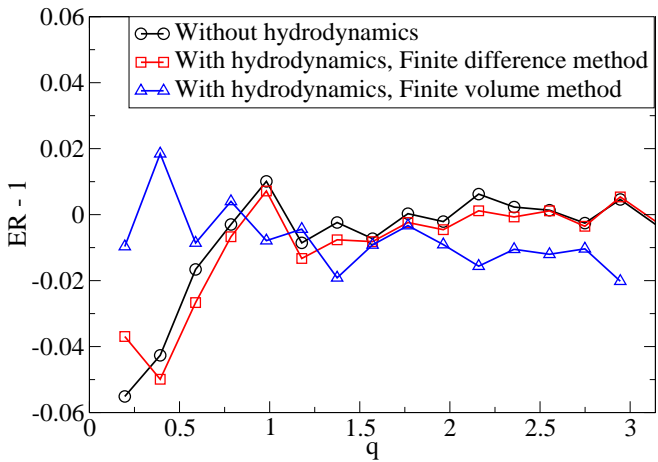}
\caption{(Color Online) Error in the equilibrium ratio as a function of wave vector magnitude, $q$, along the diagonal $q_x = q_y = q_z$ considering (1) diffusion alone and (2) coupled hydrodynamics with (i) finite difference and (ii) finite volume method for the free energy functional described by Eq. \ref{eqn:Feng} with $B=0$ for the same set of parameters in Fig. \ref{fig:test2a} and Fig. \ref{fig:test2}.}
\label{fig:test2dia}
\end{figure}

The equilibrium structure factor of a miscible binary mixture, $B=0$, (above the critical temperature) which experiences an energy cost to order parameter gradients is
\begin{equation}
 \langle|\widetilde{\psi}(\mathbf{q})|^2 \rangle = \frac{kT}{A+K\mathbf{q}^2}.
\label{eqn:test2}
\end{equation}
On a lattice, the discrete representation of the Laplacian must be accounted for, and the static spectrum reads accordingly, $ \langle|\widetilde{\psi}(\mathbf{q})|^2 \rangle = kT/(A-K L(q))$.

Since we have used the Shinozaki - Oono form for the Laplacian, Eq. \ref{eqn:solap}, to calculate $\nabla^2 \psi$ in our simulations, $-\mathbf{q}^2$ of  Eq. \ref{eqn:test2} is replaced by the Fourier transform of appropriate Laplacian $ L(\mathbf{q})$, i.e, Eq. \ref{eqn:solapft}.

Fig.  \ref{fig:test2a}  displays the simulated  $\langle|\widetilde{\psi}(\mathbf{q})|^2 \rangle$ at equilibrium on a wavenumber plane of constant $q_z$ in the absence of hydrodynamic coupling while Fig. \ref{fig:test2} shows results for the full dynamics using the two complementary  spatial discretisation approaches.  The analytical prediction  is  superimposed showing the high degree of accuracy and isotropy obtained in all situations. Only at large wave vectors   the results obtained using the finite difference method  compare better with theory than those obtained from finite volume method. We attribute this accuracy loss to the different  structure  of the lattice Laplacian  in both approaches, although the  errors are consistent with the statistical uncertainty associated to the sampling performed. To show that there is no systematic errors hidden in Fig. \ref{fig:test2a} and Fig. \ref{fig:test2}, a one dimensional plot of the  error in the equilibrium ratio is plotted against $q$, along the diagonal $q_x = q_y = q_z$ in the wave vector space in Fig. \ref{fig:test2dia}.

\begin{figure*}
\subfigure[]{\includegraphics[trim = 5mm 0mm 25mm 5mm, clip, width=0.45\linewidth]{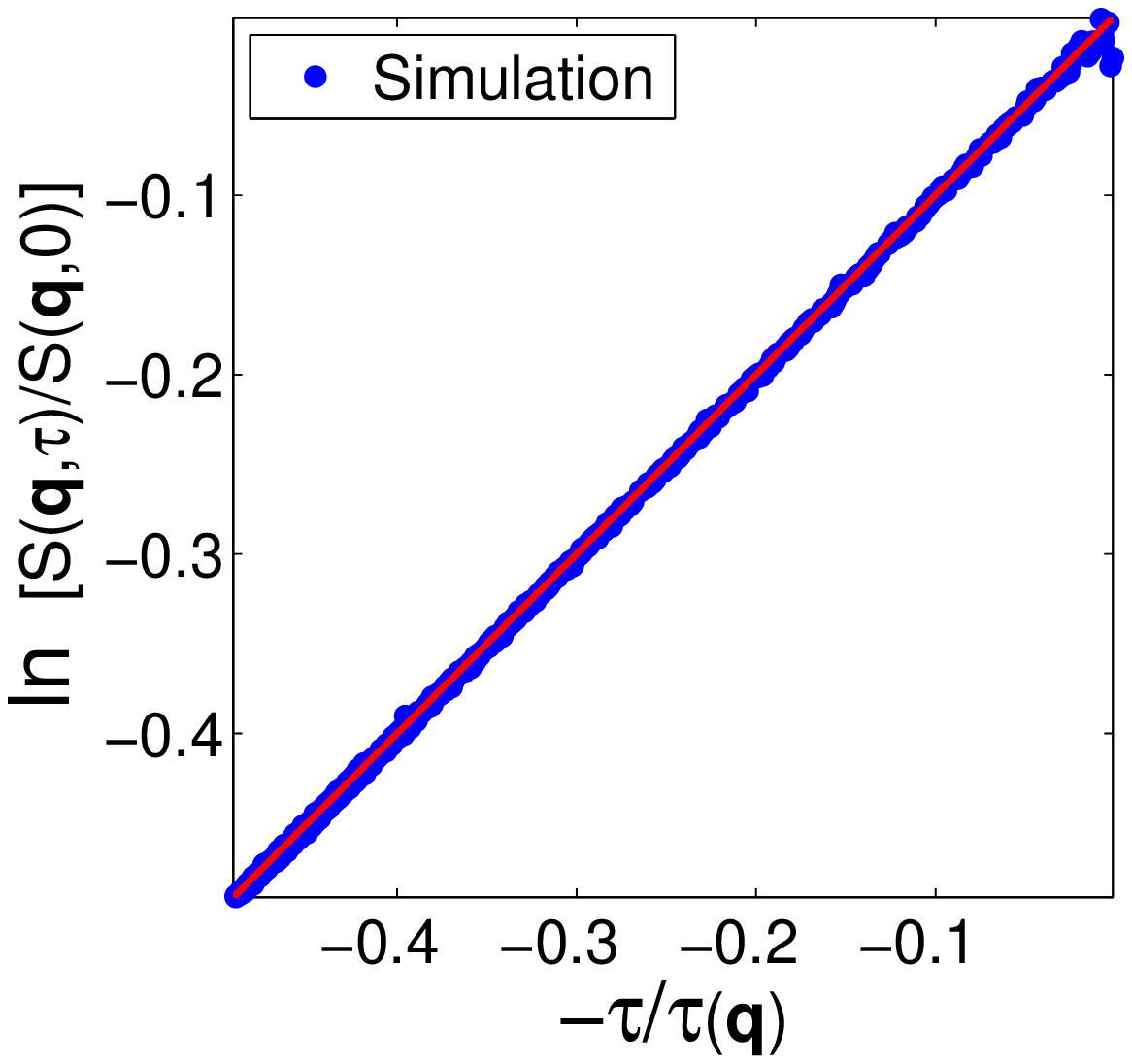}}
\subfigure[]{\includegraphics[trim = 5mm 0mm 25mm 5mm, clip, width=0.45\linewidth]{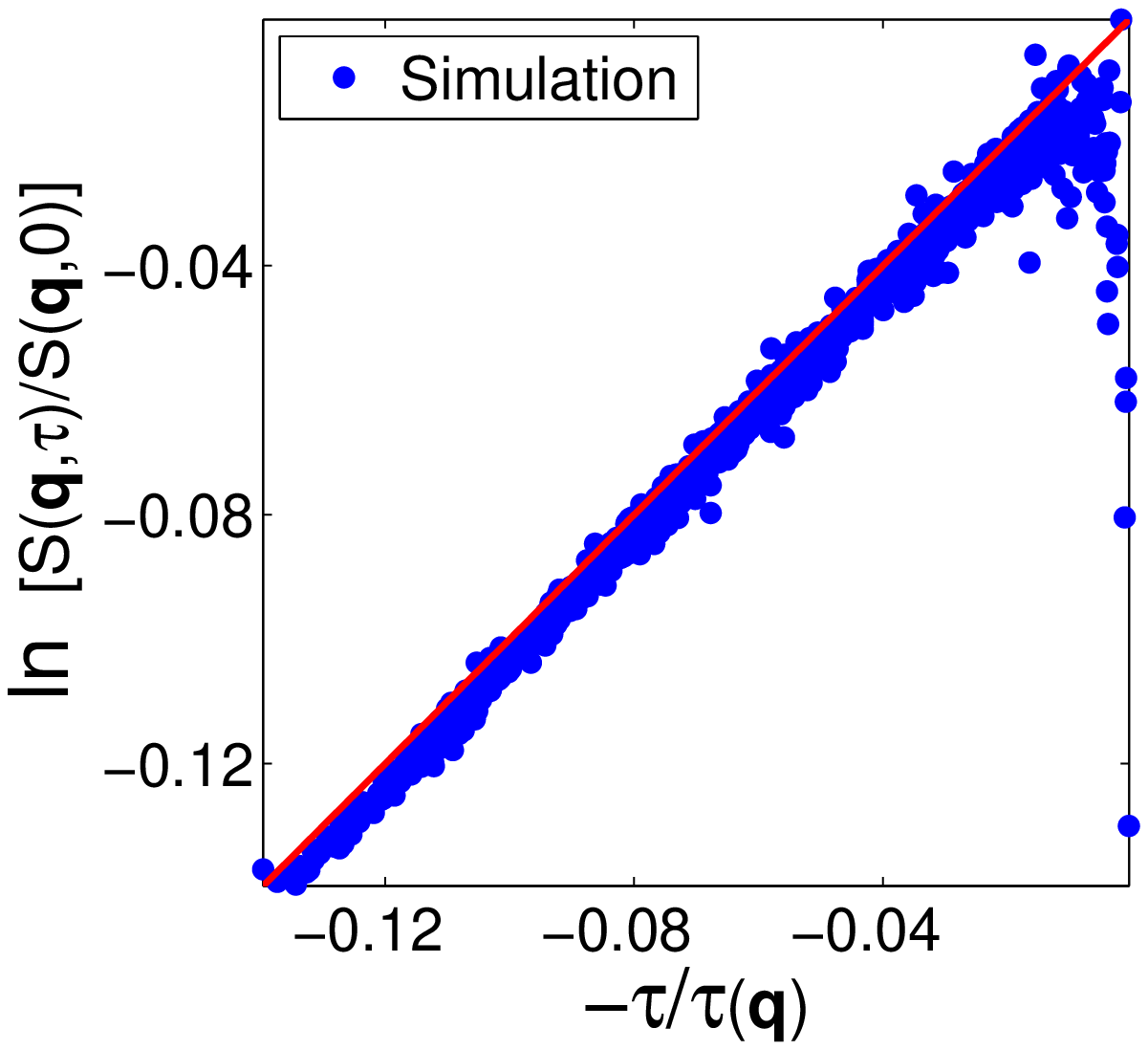}}
\caption{(Color Online) Verification of the dynamic correlation function, Eq. \ref{eqn:strf}, of the order parameter Fourier components. Finite difference scheme (a) and finite volume scheme  (b) have been used to carry out simulations on a $32 \times 32 \times 32$ lattice with $A = 0.065$, $B=0$, $K=0.04$ and $M=0.095$ with an initial equilibrium distribution. Ensemble averaging is done over $10^5$ time steps and 20 realizations.}
\label{fig:test3}
\end{figure*}

We have also analyzed the equilibrium dynamic structure factor of this miscible mixture, $S(\mathbf{q},\tau)\equiv \langle\widetilde{\psi}(\mathbf{q},t) \widetilde{\psi}(\mathbf{q},t+\tau) \rangle $, for which we have an analytic expression. Taking into account the lattice structure, it reads 
\begin{equation}
S(\mathbf{q},\tau)  = \frac{kT}{A-K L(q)}e^{-M \mathbf{q}^2 (A-K L(q))\tau}.
\label{eqn:strf}
\end{equation}
Fig. \ref{fig:test3}  displays $\ln \left[ {S(\mathbf{q},\tau)/S(\mathbf{q},0)}\right]$ as a function of the scaled time $\tau/\tau(\mathbf{q})$, where we introduce the characteristic decay time for each mode, $\tau(\mathbf{q})=  \left[ M (-L (\mathbf{q})) (A - K L(\mathbf{q})) \right]^{-1}$. The simulation results recover the expected slope with a high degree of accuracy over all the times covered for each mode for the two discretisation schemes of the fluctuating Cahn-Hilliard equation.

\subsection{Galilean invariance}

\begin{figure*}
\subfigure[]{\includegraphics[trim = 15mm 0mm 20mm 0mm, clip, width=0.45\linewidth]{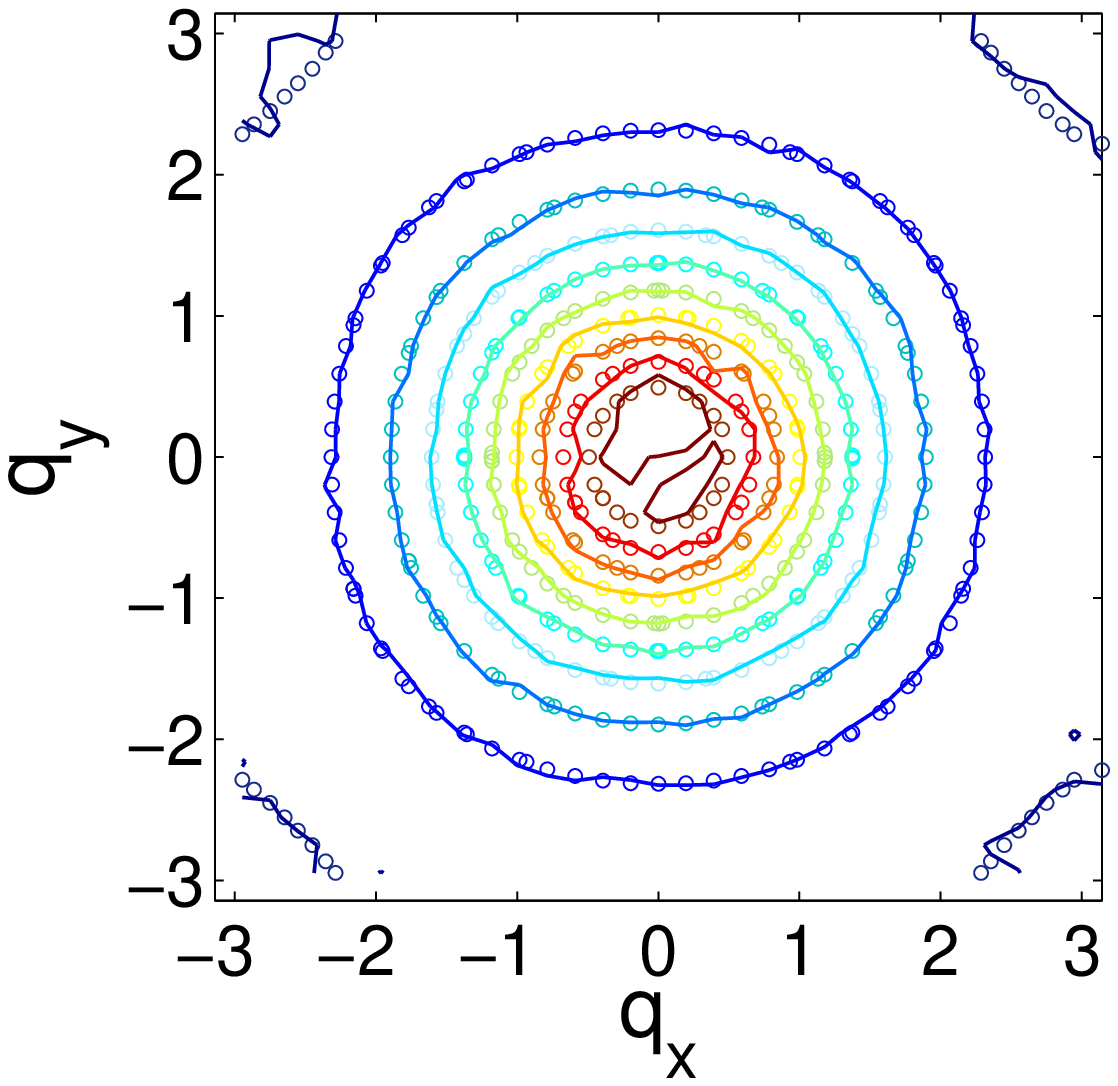}}
\subfigure[]{\includegraphics[trim = 15mm 0mm 20mm 0mm, clip, width=0.45\linewidth]{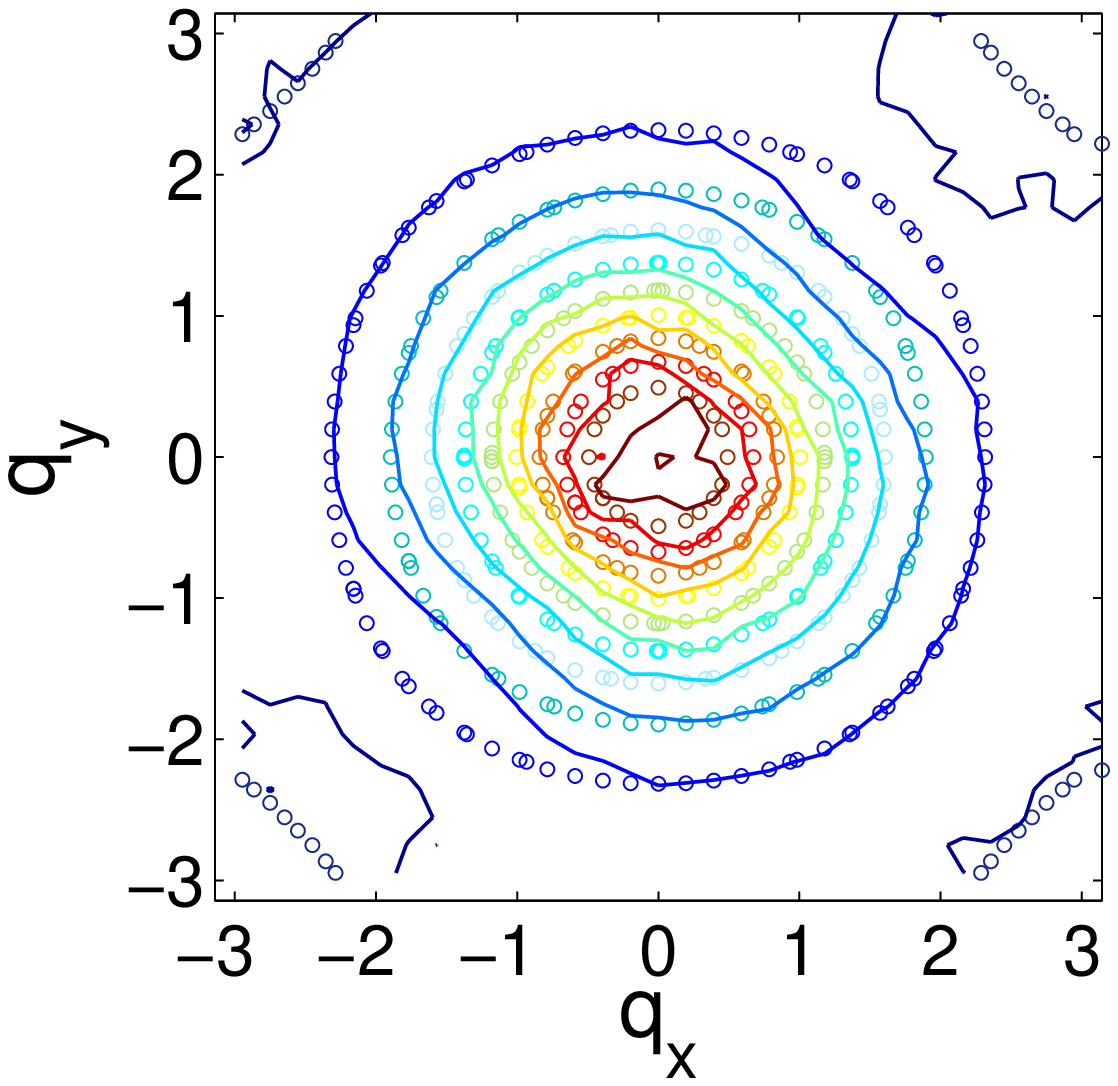}}
\caption{(Color Online) Galilean invariance of the scheme is tested by applying a uniform velocity field along a diagonal direction. Constant values of $\langle|\widetilde{\psi}({\bf q})|^2 \rangle$ from the simulations are plotted along with theoretical predictions as symbols using the same parameters as in Fig. \ref{fig:test2}. (a) At small flow velocities, $Ma=0.08$, correct equilibrium is maintained in the simulations. (b) However at large flow velocities, $Ma=0.57$, an anisotropic distribution of the order parameter fluctuations develops.}
\label{fig:galinv}
\end{figure*}

The coupling of the order parameter dynamics to the fluid motion must respect  Galilean invariance. In order to test if the proposed algorithm recovers this basic symmetry, we have imposed a constant velocity along one of the system's diagonal, $x=y$.  Fig. \ref{fig:galinv}  displays  the order parameter static structure factor, $S(\mathbf{q})=\langle|\widetilde{\psi}({\bf q})|^2 \rangle$, for a miscible mixture with an energy cost gradient, subject to a uniform flow with different magnitudes. Due to Galilean invariance, $S(\mathbf{q})$ must not be affected by the fluid motion and must coincide with the equilibrium curves  in Fig. \ref{fig:test2a} .  

At small flow rates (small $Ma$), Fig. \ref{fig:galinv}.a, we do not see any deviation from the equilibrium predictions, as expected. However, increasing the velocity for $Ma>\frac{1}{2}$, Fig. \ref{fig:galinv}.b shows the development of an anisotropic structure factor, which we attribute to the numerical dissipation associated with advection terms in the order parameter conservation equations. Although in principle, the proposed LB algorithm does not ensure Galilean invariance at high $Ma$ ( a situation which can be improved with complementary LB implementations~\cite{karlin2009}), the main source for inaccuracies comes from numerical dissipation in the order parameter dynamics. Numerically less dissipative schemes such as operator splitting may be resorted to avoid these limitations \cite{LevequeBook}.  However, in our simulations we have considered only RK algorithms, which recover the correct behavior  for small $Ma$ flows.

\subsection{Fluctuating interfaces}

\begin{figure}
\includegraphics[trim = 20mm 0mm 35mm 8mm, clip, width=0.9\linewidth]{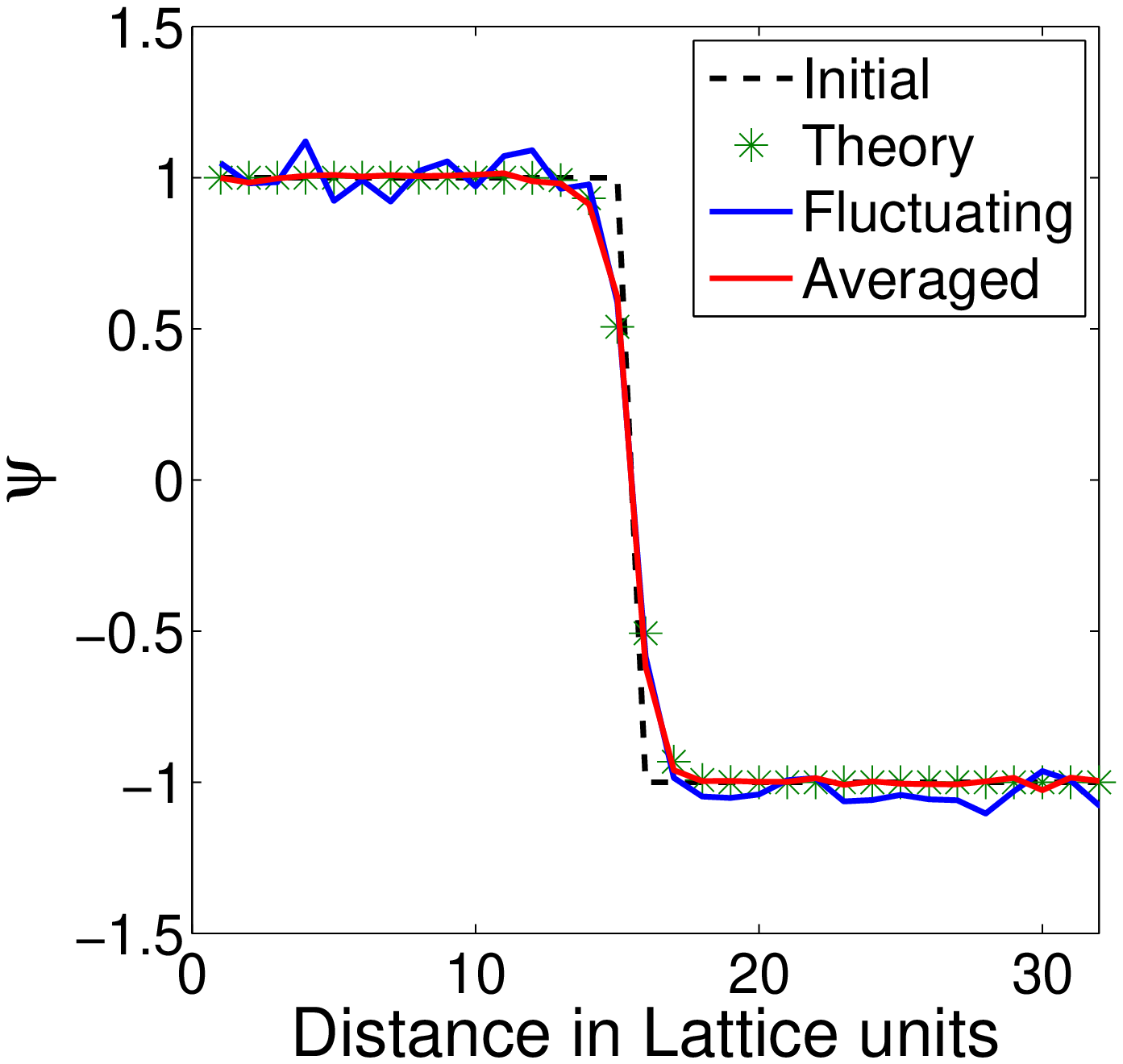}
\caption{(Color Online) Mean equilibrium profile of the order parameter for two fluid phases coexisting through a planar interface.  The dashed line is the initial set sharp profile on a 64 x 64 lattice of interface in order parameter with left-right symmetry (only left half is shown in the plot). Symbols show the theoretical predictions, (Eq. \ref{eqn:tanh}), the continuous line is an instantaneous profile from simulations while the thick line corresponds to the ensemble averaged profile.  The continuous line illustrates the magnitude of fluctuations around the mean shape. Ensemble averaging is done after attaining equilibrium($10^5$ time steps) over $4 \times 10^5$ time steps and $7$ realizations. Parameters used in the simulation are $-A=B=0.025$, $K=0.01$ and $M=0.1$.}
\label{fig:tanh}
\end{figure}

\begin{figure}
\includegraphics[trim = 0mm 0mm 0mm 0mm, clip, width=0.7\linewidth]{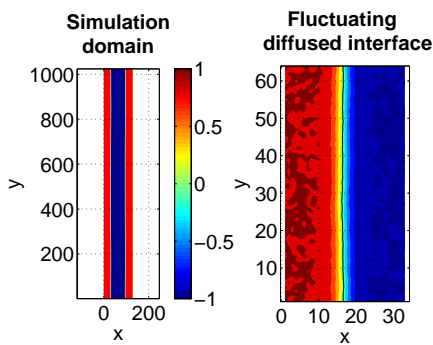}
\caption{(Color Online) Instantaneous order parameter field used for capturing the capillary spectrum. Here two fluid phases coexist through two planar fluctuating interfaces. The full domain used for simulation is shown on the left side and the fluctuating diffused interface is on the right side. The continuous line is for $\psi=0$.}
\label{fig:interface}
\end{figure}

\begin{figure}
\includegraphics[trim = 10mm 0mm 14mm 8mm, clip, width=0.9\linewidth]{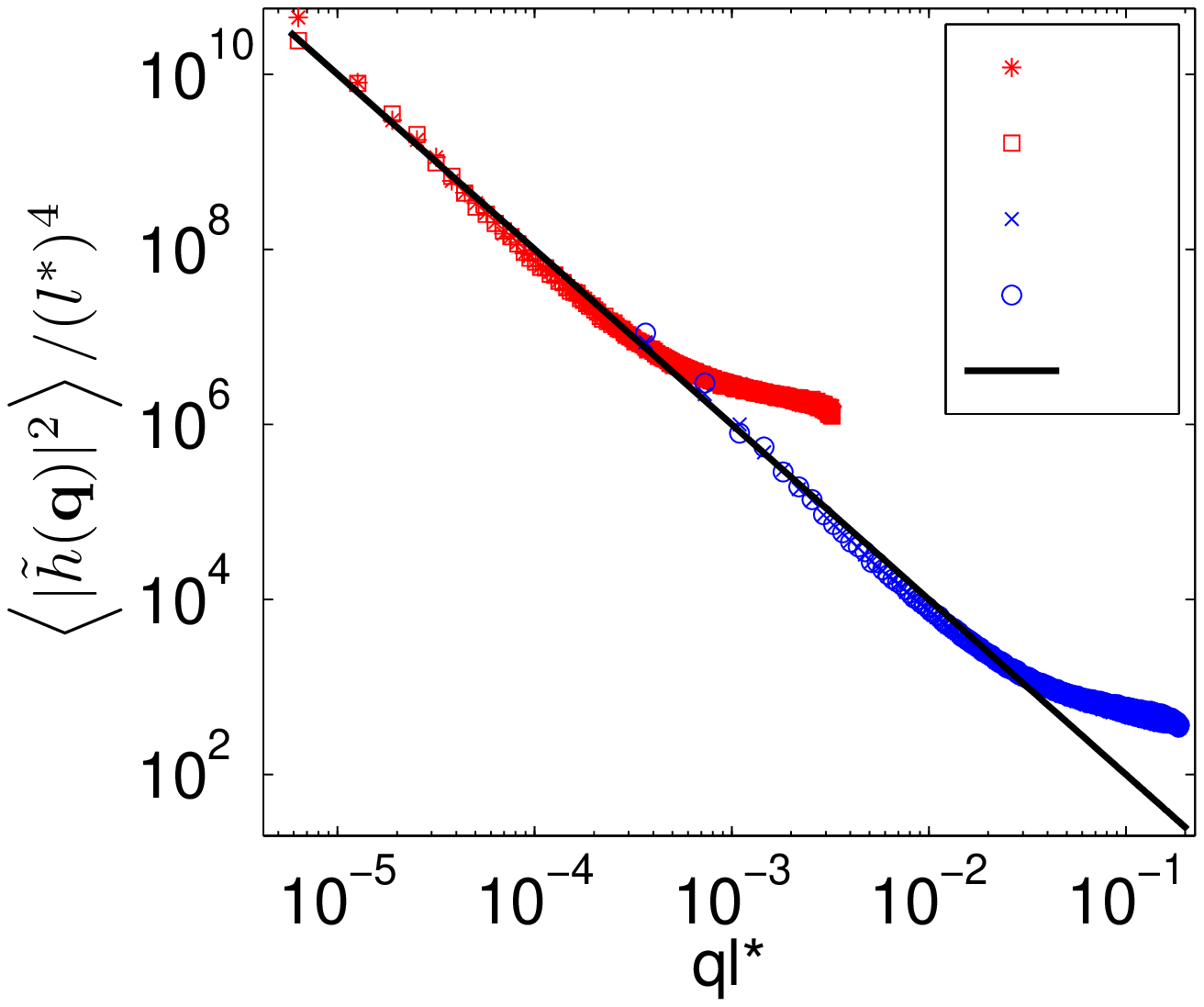}
\caption{(Color Online) Logarithmic plot of the  interfacial height fluctuation spectrum as a function of the wave vector magnitude. Symbols show the simulations results and continuous line correspond to  the theoretical prediction (Eq. \ref{eqn:flucinterf}). The wave vector magnitude is scaled with the capillary length, $l^*$, and  the magnitude of the height  fluctuations has also been scaled with $l^{-4}$ to highlight the universal nature of the capillary spectrum, which is recovered over several orders of magnitude. Four different symbols \textcolor{red}{\textasteriskcentered, \tiny$\square$,\normalsize} \textcolor{blue}{ $\times$, $\circ$} correspond to simulations with $kT=10^{-7}$ using finite-difference method, $kT=10^{-7}$ using finite-volume method, $kT=1/3000$ using finite-difference method and $kT=1/3000$ using finite-volume method respectively,  on a $1024 \times 128$ lattice (See Fig. \ref{fig:interface}). Free energy and LB  simulation parameters are $-A=B=0.05, K=0.2, M=0.1$ and $\tau = 0.45$.}
\label{fig:flucinterf}
\end{figure}
All the  tests described above have used a harmonic free energy functional.  Below, we present a test of the model including the quartic anharmonicity in the free energy. At two phase coexistence, with $A < 0$ and $B > 0$, the order parameter variation across the diffuse interface separating the two phases is the well-known hyperbolic tangent of Eq. \ref{eqn:tanh}.  In Fig. \ref{fig:tanh} we show the order parameter profile across the interface,  averaged over time and initial conditions. We have verified that the mean profile follows Eq. \ref{eqn:tanh} with a characteristic width predicted by Eq. \ref{eqn:intthick}.

Fluctuations about the mean profile are in general complicated. However, long-wavelength harmonic fluctuations are well-described by capillary wave theory  \cite{RowlinsonBook, desai1983}. The  energy of an interface with instantaneous height, $h(x,y)$ is approximated as
\begin{equation}
 \Delta F_s = \frac{1}{2} \gamma \int d^2 \mathbf{x} (\nabla h)^2
\end{equation}
In Fourier space, this is  
\begin{equation}
 \Delta F_s = \frac{\gamma}{2} \sum_{\mathbf{q}} \mathbf{q}^2 |\tilde{h}(\mathbf{q})|^2,
\end{equation}
from which it follows that
\begin{equation}
 \langle |\tilde{h}(\mathbf{q})|^2 \rangle =  \frac{kT}{\gamma \mathbf{q}^2}.
\label{eqn:flucinterf}
\end{equation}
Since our simulations evolve the entire order parameter field, which has both short-wavelength bulk fluctuations and long-wavelength capillary fluctuations, it is necessary to tune parameters appropriately to capture the capillary fluctuations. This is ensured when the thermal capillary length  $l^* = \sqrt{\frac{kT}{\gamma}}$, the interfacial width $l$, and the system size $\Lambda$ obey $l^* \ll l \ll \Lambda$. The first inequality ensures that the energy scale of the thermal fluctuations excites capillary modes and not bulk order parameter modes, while the second ensures that the long-wavelength capillary regime is accessible in the simulation. The capillary length condition is equivalently $\gamma l^2 / kT >> 1$.

We have carried out simulations on a system of size $128 \times 1024$ where interfaces of linear dimension of $1024$ are symmetrically placed about the center of the domain at a gap of $64$ lattice units (see Fig. \ref{fig:interface}) at two different temperatures. Results from these simulations are shown in Fig. \ref{fig:flucinterf} using both finite difference and finite volume methods. In diffuse interface models, alternative definitions of the interface and its location are possible \cite{Blokhuis2009}. We have used a simple linear interpolation to determine the location of the interface as the zero of the order parameter. The cross over time for roughening transition and the longest relaxation time \cite{Rothman1995, Rothman1996} may be estimated as $\sim 10^4$ and $\sim 10^3$ time steps. Therefore, simulation data was collected only after $10^5$ times steps, to ensure stationarity of the fluctuations. The logarithmic plot of Fig. \ref{fig:flucinterf} shows that the algebraic theoretical prediction can be recovered over several orders of magnitude by scaling appropriately the wave vector and height spectrum magnitudes and changing the system parameters. Exploiting the underlying scaling  structure of the interface height fluctuations, we can combine several  numerical simulations with appropriate fluid parameters to reconstruct the whole universal curve, a strategy already exploited in  the kinetics of phase-separating  fluid mixtures~\cite{Ignacio2002}. Since the quartic anharmonicity is essential in maintaining the interface and its fluctuations, this provides a non-linear test of the  equilibriation in our numerical scheme.

\section{Conclusions and outlook}
\label{sec:summary}

A hybrid method for the numerical solution of the model H equations has been developed and validated. A fluctuating lattice Boltzmann algorithm is used for hydrodynamics while a stochastic method of lines is proposed for order parameter conservation equation. Spatial discretisation in the latter case may be done using finite difference or a finite volume schemes both of which ensure correct FDT at the lattice level. FLBE takes care of fluctuations in momentum at the lattice level. The momentum and order parameter equations are coupled through stress and advection terms. The accuracy of the algorithm is demonstrated through various hydrodynamic and order parameter fluctuation tests. The capillary spectrum of height fluctuations is reproduced accurately.

There are several situations where simulations of fluctuating hydrodynamics of binary fluid system is necessary. For example, our method can be used to study phenomena such as critical fluctuations in symmetric binary mixtures and nucleation in asymmetric binary mixtures. In the light of discussions in section \ref{sec:comparison}, the role of noise in the spinodal decomposition of a binary system remains unclear \cite{Gonnella1999}. This method may be successfully employed in studying the noise driven growth in different regimes of the decomposition process. Similarly interface fluctuations play an important role in several meso scale phenomena such as fluctuations driven spreading of nano droplets on solid surfaces \cite{stone2005}, dewetting of thin films \cite{willis2009} and break up of nano jets \cite{eggers2002}. Traditionally, molecular dynamics simulations have been used to study these problems. We expect our mesoscale algorithm to be an effective complement to MD simulations which are currently limited to short time scales.

\bibliography{reference}

\begin{acknowledgments}
I. Pagonabarraga acknowledges MICINN(FIS2008 - 04386) and DURSI (2009 SGR - 634) for financial support. We thank K. Stratford, R Govindarajan and M. Gross, for the discussions and M. E. Cates for a critical reading of the manuscript. R. Adhikari acknowledges an Europa Transnational Access grant to visit Daan Frenkel in AMOLF where this work was initiated.
\end{acknowledgments}

\appendix
\section{Comparison of discrete Laplacian operators}
\label{sec:lap}

In order to ensure the isotropy of the discrete form of the Laplacian operator, we have compared four available expressions of the operator existing in the literature. Controlling the degree of isotropy of the diffusive term of the order parameter governing equation (Eq. \ref{eqn:advdiff}), i.e, $\nabla^2 \mu$ is crucial to avoid spurious interface pinning. Note that the evaluation of chemical potential ($\mu$) itself contains Laplacian of order parameter. We give the details of the comparisons here.

\begin{figure}
\includegraphics[width=0.75\linewidth]{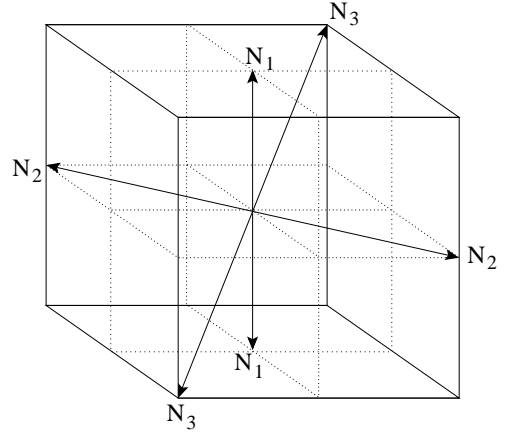}
\caption{Stencil used for Laplacian calculation for various schemes illustrated in this appendix \ref{sec:lap}. Here $N_1$ is for the nearest neighbors, $N_2$ is for next nearest neighbors and $N_3$ is for next next nearest neighbors. For clarity only one pair of each of them is marked.}
\label{fig:lapcube}
\end{figure}

Consider a 3d cubic lattice as shown in Fig. \ref{fig:lapcube}: it has 6 nearest neighbors, denoted as $N_1$, 12 next nearest neighbors, denoted as N$_2$ and 8 next next nearest neighbors, denoted as $N_3$. Correspondingly the set of lattice vectors with one, two and three non zero components form the set $\mathbf{c}_i^{N_1}$, $\mathbf{c}_i^{N_2}$ and $\mathbf{c}_i^{N_3}$ respectively where $[\mathbf{c}_i^{N_1},\mathbf{c}_i^{N_2},\mathbf{c}_i^{N_3}] \in \mathbf{c}_i$. Then
\begin{eqnarray}
[\nabla^2 \psi(\mathbf{r})]_{CD} &=& \sum_{i=1}^{6} \psi(\mathbf{r}+\mathbf{c}_i^{N_1}) - 6\psi(\mathbf{r})\\
\label{eqn:fdlap}
[\nabla^2 \psi(\mathbf{r})]_{PK} &=& \frac{14}{30} \sum_{i=1}^{6} \psi(\mathbf{r}+\mathbf{c}_i^{N_1}) + \frac{3}{30} \sum_{i=1}^{12} \psi(\mathbf{r}+\mathbf{c}_i^{N_2}) \nonumber\\
&+& \frac{1}{30} \sum_{i=1}^{8} \psi(\mathbf{r}+\mathbf{c}_i^{N_3}) - \frac{128}{30} \psi(\mathbf{r})\\
\label{eqn:pklap}
[\nabla^2 \psi(\mathbf{r})]_{SO} &=& \frac{6}{22} \sum_{i=1}^{6} \psi(\mathbf{r}+\mathbf{c}_i^{N_1}) + \frac{3}{22} \sum_{i=1}^{12} \psi(\mathbf{r}+\mathbf{c}_i^{N_2}) \nonumber 
\\&+& \frac{1}{22} \sum_{i=1}^{8} \psi(\mathbf{r}+\mathbf{c}_i^{N_3}) - \frac{80}{22} \psi(\mathbf{r})\\
\label{eqn:solap}
[\nabla^2 \psi(\mathbf{r})]_{LB} &=& \frac{1}{9} \left[ \sum_{i=1}^{26} \psi(\mathbf{r}+\mathbf{c}_i) - 26 \psi(\mathbf{r}) \right]
\label{eqn:lblap}
\end{eqnarray}
where $\psi(\mathbf{r}) =\psi(x,y,z)$. The suffixes $CD$,  $PK$, $SO$ and $LB$ stand for central difference, Patra-Kartunnen, Shinozaki-Oono and lattice Boltzmann, respectively. Eq. \ref{eqn:fdlap} is the standard central finite difference expression. Eq. \ref{eqn:pklap} has been systematically derived by imposing conditions of rotational invariance and isotropy of the operator \cite{Patra2005}. Eq. \ref{eqn:solap} is popular in the cell-dynamics and phase separation studies \cite{Shinozaki1993}. Eq. \ref{eqn:lblap} is a simple expression used in lattice Boltzmann simulations \cite{Desplat2001}.  The corresponding Fourier transforms are 
\begin{eqnarray}
&&\left[ L(\mathbf{q}) \right]_{CD} = 2 \left\{\left[c_x + c_y + c_z \right] - 3 \right\}\\
\label{eqn:fdlapft}
&&\left[ L(\mathbf{q}) \right]_{PK} = \frac{1}{30}\left\{28\left[c_x + c_y + c_z \right] \right. \nonumber\\
 && \quad \quad \left. + 12 \left[c_x c_y + c_x c_z + c_y c_z \right]  + 8 \left[ c_x c_y c_z \right] -128 \right\}\\
\label{eqn:pklapft}
&&\left[ L(\mathbf{q}) \right]_{SO} = \frac{1}{22} \left\{12\left[c_x + c_y + c_z \right] \right. \nonumber\\ 
&&\quad \quad \left. + 12 \left[c_x c_y + c_x c_z + c_y c_z \right]  + 8 \left[ c_x c_y c_z \right] -80 \right\}\\
\label{eqn:solapft}
&&\left[ L(\mathbf{q}) \right]_{LB} = \frac{1}{9}\left\{2\left[c_x + c_y + c_z \right] \right. \nonumber\\ 
&&\quad \quad \quad \left. + 4 \left[c_x c_y + c_x c_z + c_y c_z \right]  + 8 \left[ c_x c_y c_z \right] -26 \right\}
\label{eqn:lblapft}
\end{eqnarray}
where $c_x = \cos{q_x}, c_y = \cos{q_y}, c_z = \cos{q_z} $.

\begin{figure*}
\subfigure[]{\includegraphics[width=0.32\linewidth]{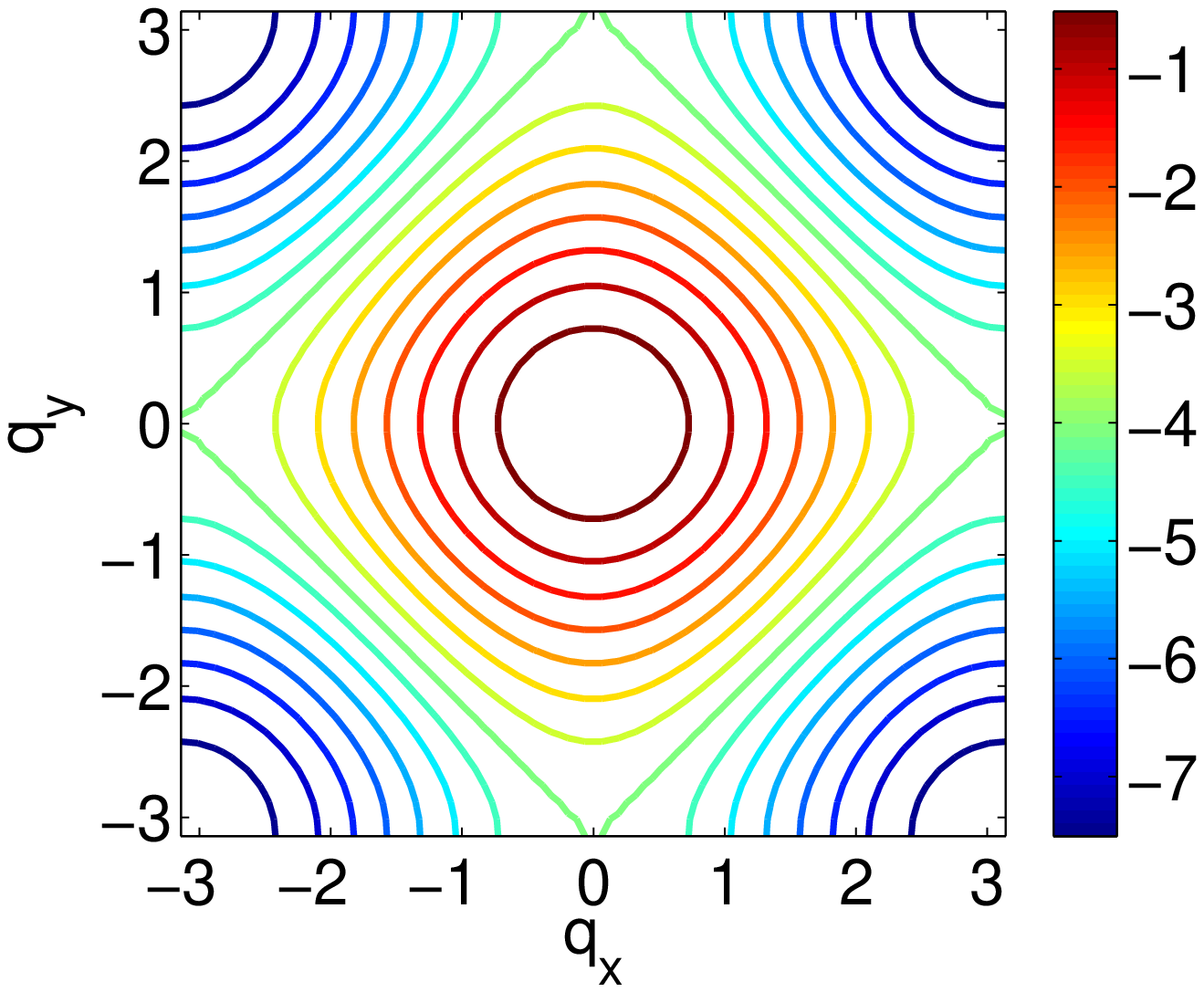}}
\subfigure[]{\includegraphics[width=0.32\linewidth]{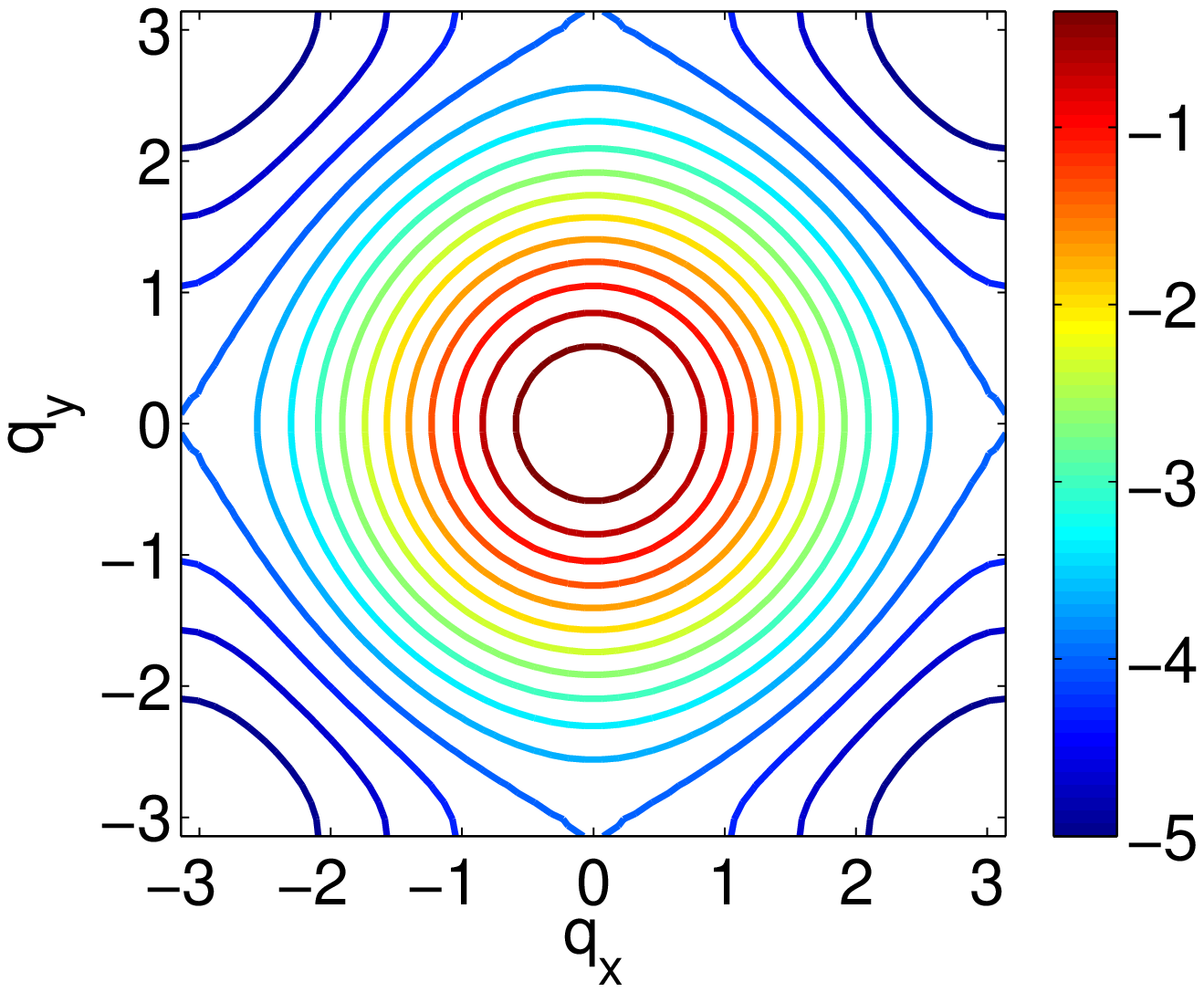}}
\subfigure[]{\includegraphics[width=0.32\linewidth]{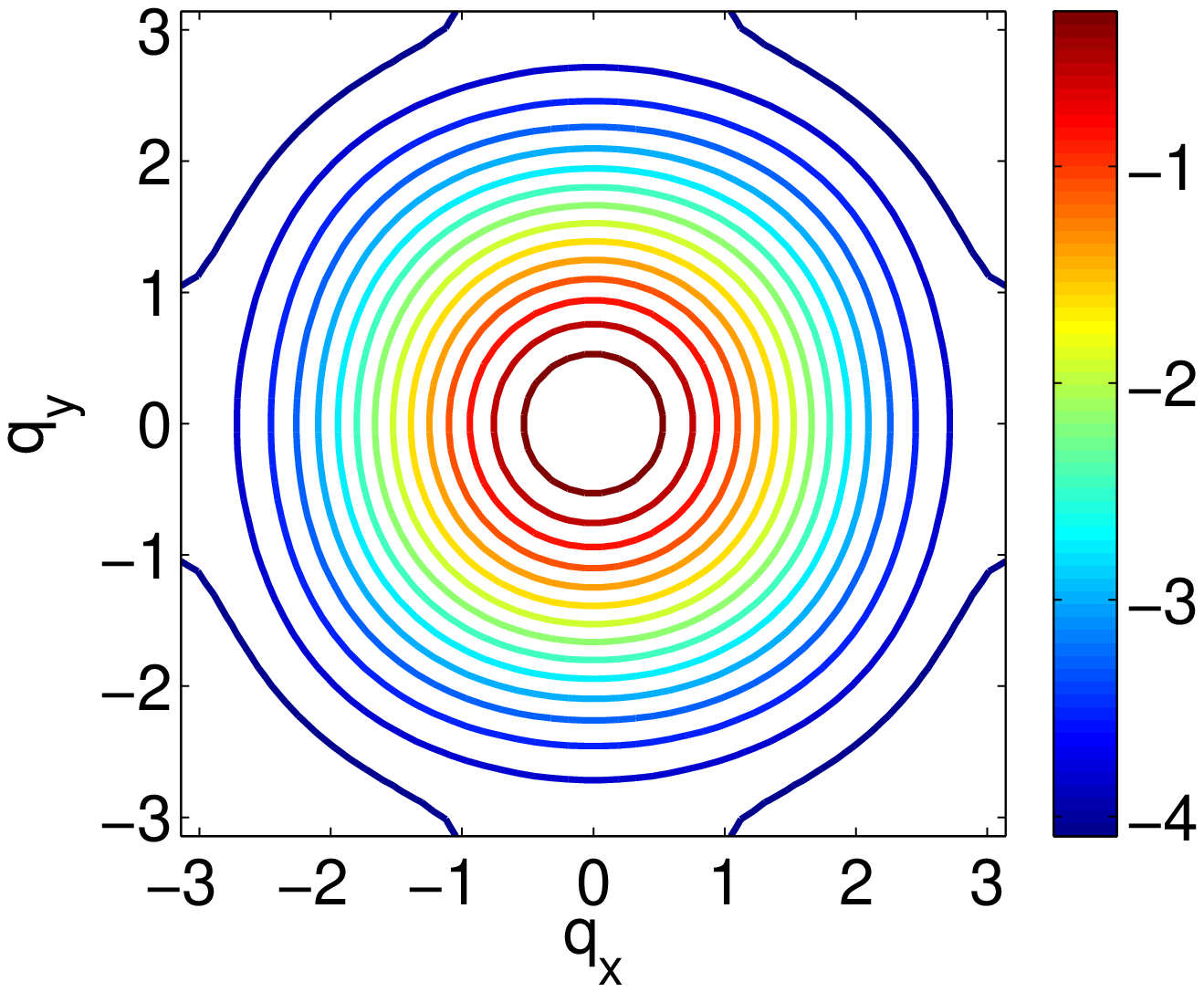}\label{fig:lapsso}} \\
\subfigure[]{\includegraphics[width=0.32\linewidth]{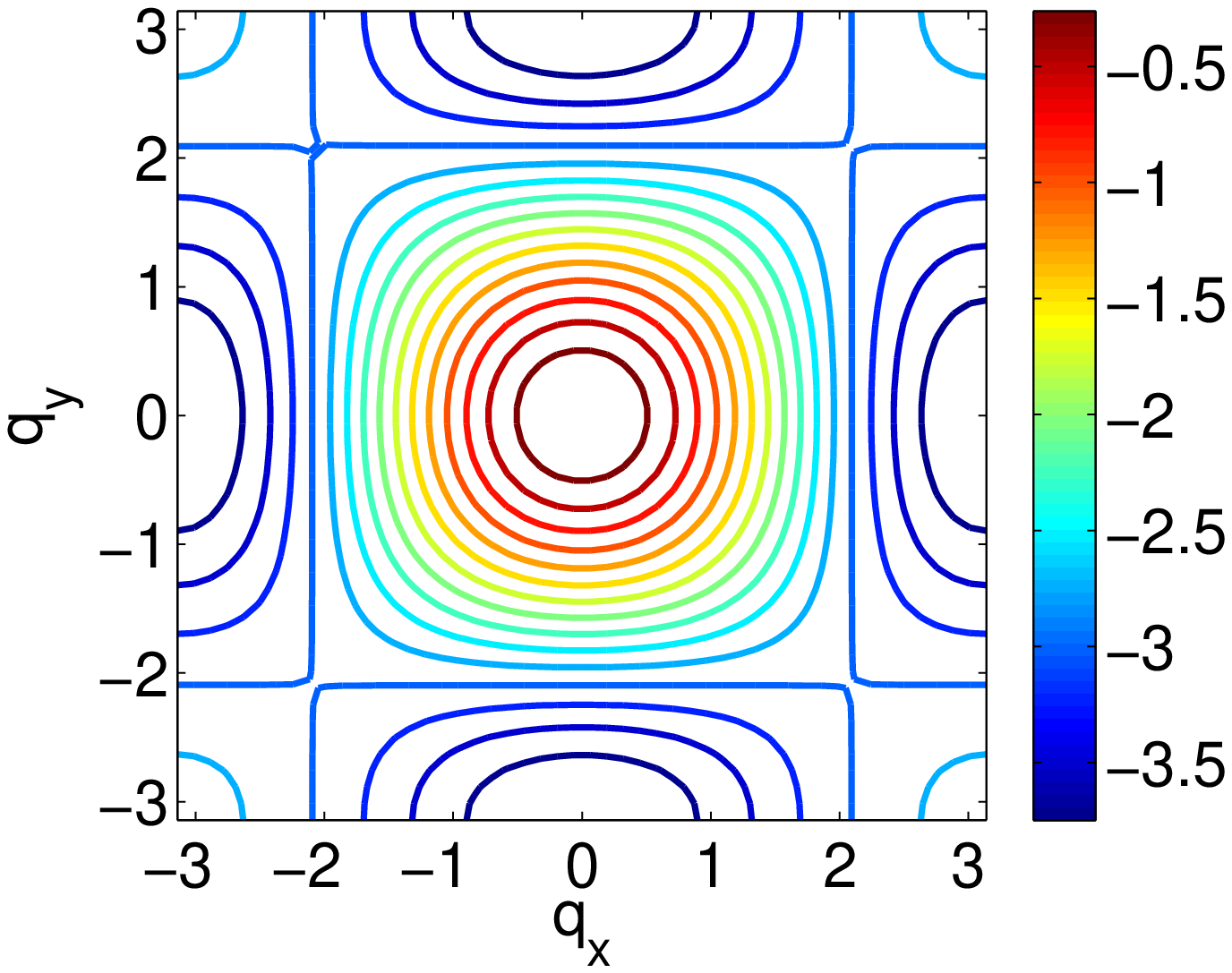}} 
\subfigure[]{\includegraphics[width=0.32\linewidth]{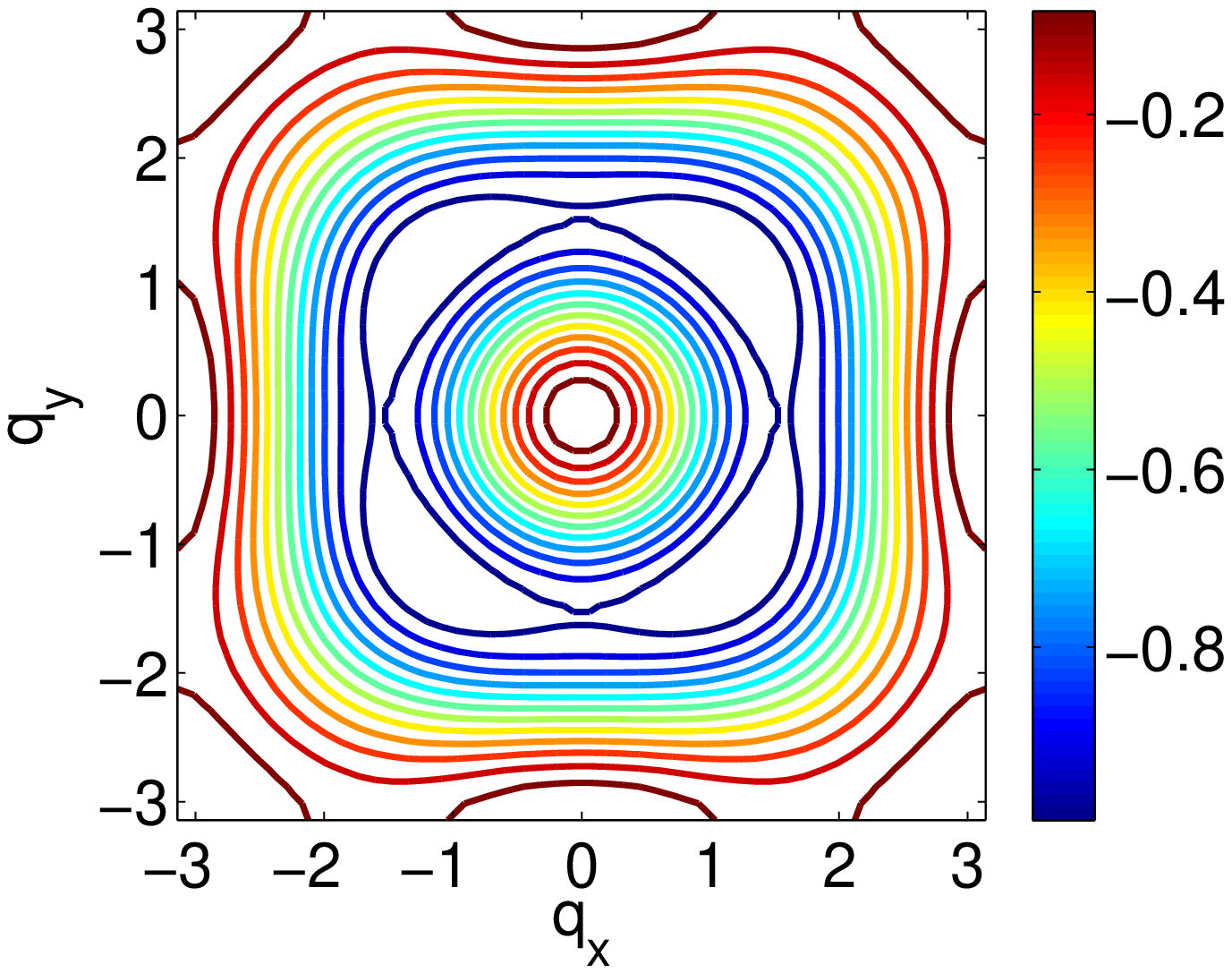}\label{fig:lapsfv}}
\caption{(Color Online) Fourier transform of Laplacian operators in $q_z=$ constant plane (a) finite difference operator, Eq. \ref{eqn:fdlapft} (b) Patra and Kartunnen Eq. \ref{eqn:pklapft} \cite{Patra2005}, (c) Shinozaki and Oono Eq. \ref{eqn:solapft} \cite{Shinozaki1993}, (d) lattice based schemes Eq. \ref{eqn:lblapft} \cite{Desplat2001}, (e) calculating divergence of flux defined on links in the finite volume approach (Eq. \ref{eqn:fvlapft} in \ref{sec:FV})}
\label{fig:laps}
\end{figure*}

Clearly all the Laplacian operators are negative definite (except at $q=0$). Fig. \ref{fig:laps} shows the magnitude of the different expressions of the Laplacian operator in  wave vector planes with constant $q_z$. These plots clearly display the four fold symmetry of the lattice. Nonetheless, the effect is less pronounced for the expression suggested by Shinozaki and Oono (\ref{fig:lapsso}) and so we have used Eq. \ref{eqn:solap} for the calculations in section \ref{sec:smol} - \ref{sec:results}.

In the finite volume approach to solve  the order parameter evolution, fluxes are calculated (Eq. \ref{eqn:fvgrad}) on the links connecting the lattice nodes. Ensuring FDT leads to an equivalent Laplacian operator whose Fourier transform reads
\begin{eqnarray}
\left[ L(\mathbf{q}) \right]_{FV} &=& - \frac{1}{9}\left\{\left[2s_x + s_x c_y c_z \right]^2 + \left[2s_y + s_y c_x c_z \right]^2 \right.  \nonumber\\
&&\left.  + \left[2s_z + s_z c_x c_y \right]^2 \right\}
\label{eqn:fvlapft}
\end{eqnarray}
where $s_x = \sin{q_x}, s_y = \sin{q_y}, s_z = \sin{q_z}$. This is also plotted in Fig. \ref{fig:laps} for comparison purpose.

\section{Fluctuating Navier-Stokes solver}
\label{sec:flbeappen}

In this appendix we describe FLBE method we use to solve  the fluctuating Navier-Stokes equations. A self-contained derivation is given below to include noise \cite{adhikari2005} and external force densities \cite{nash2008} to the standard LB model.

In a standard $DdQn$ LBE model where the velocity is discretized into $n$ components in a $d$ dimensional space, the discrete form of the fluctuating Boltzmann equation  is given  in Eq. \ref{eqn:flbe}, where the moments of the single particle distribution function $f_i$,  are expressed in Eq. \ref{eqn:momreln}. A multi-scale expansion, or a moment closure method, shows that the above equation has Eq. \ref{eqn:ns} as its hydrodynamic limit \cite{SucciBook}. Since FLBE is a  hyperbolic equation with local non-linearities, it is considerably easier to solve than Eq. \ref{eqn:ns}, which has a parabolic-hyperbolic character with advective non-linearities. The methodology of the FLBE has been explained in detail in \cite{dunwegladd2007}, while the method by which force densities are added is given in detail in \cite{nash2008}. Here we outline the integration scheme we use when force densities and fluctuating forces are combined in the FLBE. 

We can rearrange equation, Eq. \ref{eqn:flbe} to obtain
\begin{equation}
 \partial_t f_i + \mathbf{c}_i. \mathbf{\nabla} f_i = R_i(\mathbf{x},t)
\label{eqn:derv1}
\end{equation}
where $R_i(\mathbf{x},t) = - \sum_j L_{ij}( f_j -f_j^0)+ \Phi_i$ represents the effects of collision, forcing and thermal fluctuations. $\Phi_i \equiv  \zeta_i -\mathbf{F}\cdot \nabla_c f$ accounts for the fluctuating and external forces acting on the distribution function. Using the method of characteristics, this set of first order hyperbolic equations can be integrated over a time interval $\Delta t$ to get
\begin{equation}
f_i(\mathbf{x}+\mathbf{c}_i \Delta t, t+\Delta t) - f_i(\mathbf{x},t) = \int_{0}^{\Delta t} ds R_i(\mathbf{x}+\mathbf{c}_is, t+s)
\end{equation}
The integral above may be approximated to second order accuracy using the trapezium rule and the resulting terms transposed to give a set of implicit equations for the $f_i$ :
\begin{eqnarray}
 f_i(\mathbf{x}+\mathbf{c}_i \Delta t, t+ \Delta t) &-& \frac{\Delta t}{2} R_i(\mathbf{x}+\mathbf{c}_i \Delta t, t+\Delta t) = \nonumber \\
 f_i(\mathbf{x},t) &-& \frac{\Delta t}{2} R_i(\mathbf{x},t) + \Delta t R_i(\mathbf{x},t) .
\end{eqnarray}
In terms of the auxiliary distribution function, Eq. \ref{eqn:barreln}, the evolution equation reduces to   
\begin{equation}
 \bar{f}_i(\mathbf{x}+\mathbf{c}_i \Delta t, t+\Delta t) = \bar{f}_i(\mathbf{x},t) + R_i(\mathbf{x},t) \Delta t .
\label{eqn:bardisc}
\end{equation}
indicating  that we can understand  LBE evolution through a simple relaxational step in which the distributions $\bar{f}_i$ are relaxed to their postcollisional values $\bar{f}_i(\mathbf{x},t^*)$,
\begin{equation}
 \bar{f}_i(\mathbf{x},t^*) = \bar{f}_i(\mathbf{x},t) + R_i(\mathbf{x},t) \Delta t,
\end{equation}
followed by a propagation step in which the postcollisional distributions are propagated along a Lagrangian trajectory without further change,
\begin{equation}
 \bar{f}_i(\mathbf{x}+\mathbf{c}_i \Delta t, t+\Delta t) = \bar{f}_i(\mathbf{x},t^*).
\end{equation}
Thus the computational part of the method is most naturally framed in terms of the auxiliary distributions $\bar{f}_i$ instead of the physical distribution functions $f_i$ themselves. To obtain the postcollisional $\bar{f}_i$ without having to refer to the $f_i$, the latter must be eliminated from Eq. \ref{eqn:bardisc}. Inverting the equations defining the $\bar{f}_i$ in Eq. \ref{eqn:barreln}, we obtain
\begin{equation}
 R_i = \left( 1+\frac{\Delta t}{2} L \right)_{ij}^{-1} [-L_{jk}(\bar{f}_k - f_k^0) + \Phi_j(\mathbf{x},t)] .
\end{equation}
Combining this with Eq. \ref{eqn:bardisc} we obtain a numerical scheme for the discrete Boltzmann equation with a general collision operator in terms of the $\bar{f}_i$:
\begin{eqnarray} 
 \bar{f}_i(\mathbf{x}+\mathbf{c}_i \Delta t, t+\Delta t) = \bar{f}_i(\mathbf{x},t) + \nonumber \\
 \left( 1+\frac{\Delta t}{2} L \right)_{ij}^{-1} [-L_{jk}(\bar{f}_k - f_k^0) + \Phi_j(\mathbf{x},t)] \Delta t.
\label{eqn:barlin}
\end{eqnarray}
For a single time relaxation operator, where $L_{ij} = \delta_{ij}/\tau$, this takes on a particularly simple form,
\begin{eqnarray}
 \bar{f}_i(\mathbf{x}+\mathbf{c}_i \Delta t, t+\Delta t) = \bar{f}_i(\mathbf{x},t) +\nonumber \\
   \frac{\Delta t}{\tau + \Delta t/2} [-(\bar{f}_i - \bar{f}_i^0)+\tau \Phi_i(\mathbf{x},t)],
\end{eqnarray}
For a nondiagonal collision operator, the collision term is best evaluated in the moment basis. For example, using a collision operator in which the ghost modes are projected out \cite{SucciBook} and the stress modes relax at a rate of $\tau^{-1}$, the post collisional $\bar{f}_i$  are given by 
\begin{equation}
 \bar{f}_i(\mathbf{x},t^*) = w_i \left( \rho + \frac{A_{\alpha} c_{i \alpha}}{c_s^{2}} + \frac{B_{\alpha \beta} Q_{i \alpha \beta}}{2c_s^4} \right)
\end{equation}
where the normalized weights $w_i$ ensure the isotropy, and $A_{\alpha}$, the momentum component of the postcollisional auxiliary distributions, is
\begin{equation}
 A_{\alpha} = \sum_{i=0}^{n} \bar{f}_ic_{i \alpha} + \rho F_{\alpha} \Delta t
\end{equation}
while $B_{\alpha \beta}$, the stress component, reads 
\begin{eqnarray}
 B_{\alpha \beta} = \sum_{i=0}^{n} \bar{f}_i Q_{i \alpha \beta} + \frac{\Delta t}{\tau + \Delta t/2} 
\left( -\sum_{i=0}^{n} \bar{f}_i Q_{i \alpha \beta} \right. \nonumber \\
\left .+ \rho v_{\alpha} v_{\beta} +
\tau(v_{\alpha}F_{\beta}+F_{\alpha}v_{\beta})+ \tau \sum_{i=0}^{n}\zeta_i Q_{i\alpha\beta} \right).
\label{eqn:bab}
\end{eqnarray}
The mass and momentum densities are obtained as $\rho = \sum_{i=0}^{n} \bar{f}_i$ and $\rho v_{\alpha} = \sum_{i=0}^{n} \bar{f}_i c_{i\alpha} + \rho F_{\alpha} \frac{\Delta t}{2}$, respectively.  The equilibria can be reconstructed from $\rho$ and $\rho \mathbf{v}$. 

\end{document}